\documentclass[11pt,oneside]{article}
\usepackage{aapreprint}
\usepackage{graphicx,wrapfig,float,slashed}
\usepackage{amsmath,amssymb,dsfont,epsfig,graphicx}
\usepackage{epstopdf}
\usepackage{ragged2e}
\usepackage{enumitem}
\usepackage{arydshln}
\usepackage[utf8]{inputenc}
\usepackage{epstopdf}
\usepackage[usenames,dvipsnames,table]{xcolor}
\usepackage{tabu}
\usepackage[toc]{appendix}
\usepackage[normalem]{ulem}
\usepackage[font=small,skip=1pt]{caption}
\allowdisplaybreaks
\usepackage{pifont}
\usepackage{subcaption}
\usepackage{cancel}
\usepackage{comment}

\usepackage{multicol}
\usepackage{geometry}
\usepackage{xcolor}

\newcommand{\nc}{\newcommand}
\nc{\non}{\nonumber}
\nc{\hc}{\hbox {h.c.}}
\nc{\noi}{\noindent}
\nc{\barx}{\bar{x}}
\nc{\pbarn}{\;\hbox {pb}}
\nc{\fbarn}{\;\hbox {fb}}

\nc{\hsp}{\hspace{0.5cm}}
\nc{\lsp}{\hspace{1cm}}
\nc{\Lsp}{\hspace{2cm}}
\nc{\LLsp}{\lsp\lsp}
\nc{\lra}{\longrightarrow}
\nc{\p}{\prime}
\nc{\sgn}{\text{sgn}}
\nc{\ph}{\varphi}
\nc{\op}{{\cal O}}
\nc{\eq}{\text{Eq.~}}

\nc{\beq}{\begin{equation}}  \nc{\eeq}{\end{equation}}
\nc{\bs}{\begin{split}}  \nc{\es}{\end{split}}
\nc{\bea}{\begin{eqnarray}}  \nc{\eea}{\end{eqnarray}}
\nc{\baa}{\begin{array}}     \nc{\eaa}{\end{array}}
\nc{\bit}{\begin{itemize}}   \nc{\eit}{\end{itemize}}
\nc{\ben}{\begin{enumerate}} \nc{\een}{\end{enumerate}}
\nc{\bce}{\begin{center}}    \nc{\ece}{\end{center}}
\nc{\bpm}{\begin{pmatrix}}   \nc{\epm}{\end{pmatrix}}
\nc{\bvt}{\begin{verbatim}}  \nc{\evt}{\end{verbatim}}

\def\gev{\;\hbox{GeV}}
\def\tev{\;\hbox{TeV}}

\def\mpl{M_{\text{Pl}}}
\definecolor{agray}{rgb}{0.95, 0.95, 0.99}
\def\mphi {m_{\phi}}

\def\mkk{m_{\rm KK}}
\def\Z2{\mathbb{Z}_2}
\def\cF{{\cal F}}

\nc{\cw}{\cos\theta_{\textsc w}}
\nc{\sw}{\sin\theta_{\textsc w}}
\nc{\cwsq}{\cos^2\theta_{\textsc w}}
\nc{\swsq}{\sin^2\theta_{\textsc w}}
\def\ir{\text{\tiny IR}}
\def\uv{\text{\tiny UV}}
\setlist[itemize]{noitemsep, topsep=0.25em}

\begin{document}
\title{\bf \!\!Dilaton portal in strongly~interacting~twin~Higgs models\!\!}

\author[1]{Aqeel Ahmed,} 
\author[2]{Barry M. Dillon,}
\author[1]{and Saereh Najjari}

\affiliation[1]{Theoretische Natuurkunde \& IIHE, Vrije Universiteit Brussel, Pleinlaan 2, 1050 Brussels, Belgium}
\affiliation[2]{Jo\u{z}ef Stefan Institute, Jamova 39, 1000 Ljubljana, Slovenia}
\emailAdd{aqeel.ahmed@vub.be}
\emailAdd{barry.dillon@ijs.si}
\emailAdd{saereh.najjari@vub.be}
\abstract{\!
We consider a  strongly interacting twin Higgs (SITH) model where an ultraviolet completion of twin Higgs mechanism is realized by a strongly coupled approximately scale invariant theory. 
Besides the Standard Model (SM) and twin sectors, the low energy effective theory contains a relatively light scalar called a dilaton --- the pseudo Goldstone boson of spontaneously broken scale invariance. 
The dilaton provides a unique portal between the SM and twin sectors whose phenomenology could provide an important probe of the twin Higgs mechanism.
As a concrete example, we consider a holographic twin Higgs model where the role of the dilaton is played by the radion. 
The phenomenology of this model is fully determined by a few parameters and our analysis concludes that at the HL-LHC ($14\tev$) and HE-LHC ($27\tev$) with $3000\!/\!{\rm fb}$ most of the natural parameter space can be probed. 
}
\keywords{Beyond the Standard Model, Neutral Naturalness, Twin Higgs, Composite Higgs, Scale Invariance, Dilaton/Radion Phenomenology \vspace{17pt}}

\toccontinuoustrue

\maketitle

\section{Introduction}
\label{Introduction}
The discovery of the SM-like Higgs boson with a mass of $125\gev$ at the Large Hadron Collider (LHC) ~\cite{Aad:2012tfa,Chatrchyan:2012xdj} has initiated an intense effort in the theoretical physics community to understand the origin of electroweak symmetry breaking (EWSB).
One of the outstanding issues which we hope to shed light on is the gauge-hierarchy problem.
Most of the beyond the SM (BSM) scenarios addressing this issue typically predict properties of the Higgs boson which are modified with respect to the SM Higgs boson, as well as a plethora of new physics states at the TeV scale.
Probing the structure of EWSB therefore requires simultaneous considerations of modified Higgs properties and bounds on new physics states within a given model. 
In recent years BSM scenarios have come under increasing tension with the ATLAS and CMS collaborations having made significant progress in probing the couplings of the Higgs boson to the SM fields, with all major production modes now observed at more than 5$\sigma$ significance, and pushing the lower bounds on many new physics states to the TeV scale. 

A particularly intuitive explanation of electroweak symmetry breaking is present in the composite Higgs paradigm~\cite{Kaplan:1983fs,Kaplan:1983sm,Georgi:1984af,Agashe:2004rs,Agashe:2006at,Contino:2006qr}.
These models employ a strongly coupled gauge theory such that confinement at the scale $\Lambda\!\sim\!\op(5\!-\!10)\tev$ triggers the spontaneous breaking of a global flavor symmetry $\mathcal{G}$ to a subgroup $\mathcal{H}$. 
The resulting Goldstone bosons transforming in the $\mathcal{G}/\mathcal{H}$ coset include the SM Higgs doublet (for review see e.g.~\cite{Contino:2010rs,Bellazzini:2014yua,Panico:2015jxa}). 
This framework provides an intriguing possibility to solve the hierarchy problem, however at the cost of introducing new states due to strong dynamics close to electroweak scale.
This runs into several phenomenological problems. 
The experimental results from direct searches for top-partners at the LHC~\cite{Sirunyan:2019xeh} and indirect constraints from EWPT and flavor physics~\cite{Ahmed:2019zxm} place stringent constraints on these states calling into question the naturalness of the composite Higgs solution to the hierarchy problem, and thus poses the {\it little hierarchy} problem.

A novel solution to this little hierarchy problem is provided for in the so-called {\it neutral naturalness} paradigm of which the twin Higgs (TH) model~\cite{Chacko:2005pe,Barbieri:2005ri,Chacko:2005vw} is the primary example. 
The twin Higgs mechanism relies on two important ingredients: (a) the SM Higgs doublet is a pseudo-Goldstone boson multiplet which emerges due to the spontaneous breaking of a global symmetry at a scale $f$, and (b) a $\mathbb{Z}_2$ exchange symmetry between the SM and a new `twin SM' which leads to the cancellation of the leading quadratic divergences in the Higgs potential through states that are not charged under SM gauge symmetry, i.e. neutral top-partners. 
In its infancy the model assumed that each SM field has a corresponding twin state, this is now referred to as the `mirror twin Higgs' (MTH) model.
Since then models have been studied which, despite being more minimal in the field content, employ the same mechanism and solve or alleviate the top-partner problem~\cite{Craig:2015pha,Craig:2016kue,Serra:2017poj,Csaki:2017jby,Xu:2018ofw}. These twin Higgs models are however only valid up to scales of $\op({5\!-\!10}) \tev$ where they require a UV completion. 
There are a number of proposals for UV completions of twin Higgs models, the most compelling of which are the composite/holographic twin Higgs models~\cite{Batra:2008jy,Geller:2014kta,Barbieri:2015lqa,Low:2015nqa,Dillon:2018wye,Xu:2019xuo}.
In the composite scenarios the twin Higgs mechanism arises as an effective description of some gauge theory which becomes strongly interacting near the TeV scale, thus we refer to this as the strongly interacting twin Higgs (SITH) model.

It is the aim of this paper to study the SITH model where the gauge theory in the UV is approximately scale-invariant.
Below the mass scale $m_\ast$ associated with the composite states charged under the SM and twin sectors, these models are described by the SM and its mirror copy which are connected through a Higgs portal. 
Due to the twin Higgs mechanism the strong dynamics scale $m_\ast$ can naturally be above the reach of the LHC without introducing fine tuning in the SM Higgs potential~\cite{Geller:2014kta,Barbieri:2015lqa,Low:2015nqa}. 
A modest choice of $m_\ast\!\sim\!\op(5)\tev$ can put the new states associated with the strong dynamics out of reach at the LHC. 
However, the flavor and EWPT constraints still require much higher scales\,\footnote{The SITH models are already helping with the flavor constraints when compared to the conventional composite Higgs models for the same level of fine tuning~\cite{Csaki:2015gfd}. This is due to the fact that the masses of the new states that are problematic in flavor physics are pushed up by a factor $\sim g_\ast\equiv\mkk/f$ without introducing extra fine tuning.}, which can not be pushed arbitrarily high for fixed value of $f$ due to perturbativity constraints, i.e. $m_\ast\!\lesssim \sqrt{2}\,\pi f$~\cite{Contino:2017moj}. 
However, if the new strong dynamics is conformal in the UV then the flavor scale can be well separated from the twin Higgs symmetry breaking scale $f$. 
This scenario is motived by the proposal of alleviating flavor problem in technicolor models~\cite{Luty:2004ye}.  
Alternative possibility is that the SM and twin sector fermions acquire mass through mixings with heavy composite states, and hierarchical couplings in the flavour sector are explained through the renormalization group evolution (RGE) of the couplings from the UV~\cite{Kaplan:1991dc}. 
In this scenario the top quarks of both the SM and twin sectors are mostly composite, whereas the light fermions are mostly elementary. The gauge bosons of the twin Higgs model are external to the conformal field theory (CFT), and the SM Higgs boson in this scenario is a composite pseudo-Goldstone state. 

The approximate scale invariance in the SITH model is broken spontaneously near the scale $f$ at which the global symmetry~$\mathcal{G}$ is spontaneously broken.
In the low energy effective field theory (EFT), besides the SM and its twin copy, we will thus have:
\bit\itemsep0em
\item[(i)] a relatively light scalar, the dilaton; this is the pseudo-Goldstone boson associated with the spontaneous breaking of scale invariance,
\item[(ii)] a new portal between the SM and twin sectors, mediated by the dilaton.
\eit
The scale invariance is broken spontaneously at scale $\Lambda_\cF\!=\!4\pi\cF$ by confinement in the strongly coupled theory, where in low energy the symmetry is realized nonlinearly by the dilaton with vacuum expectation value (VEV) denoted by $\cF$~\cite{Salam:1970qk}.
The couplings of the dilaton to the SM and twin sector states can be determined purely by symmetry considerations. More specifically, in the infrared (IR) these couplings can be calculated in terms of:~(a) the explicit breaking of scale invariance in the UV, and (b) the mass dimensions of the operators in the IR EFT.
The phenomenology of the dilaton and the more formal aspects related to the dilaton potential have received a lot of attention in recent years~\cite{Goldberger:2008zz,CPR:2010,Chacko:2012sy,Bellazzini:2013fga,Coradeschi:2013gda,Appelquist:2019lgk}.
Of most relevance to this work are the studies of dilaton phenomenology in composite SM extensions~\cite{Fan:2008jk,Vecchi:2010gj,Chacko:2013dra,Chacko:2014pqa}.
What makes this an interesting consideration is that the dilaton acts as a portal between the SM and the twin sectors and, unlike the Higgs portal, it couples to the both sectors with similar strength\,\footnote{The scalar, fermion and vector portals between the SM and twin sectors are studied in~\cite{Bishara:2018sgl,Chacko:2019jgi}, where the scalar portal is induced through its mass mixing with the Higgs, which is very different than a dilaton portal.}.

As a tool in deriving predictions from such a construction, we will use a holographic 5D set-up, where the AdS/CFT correspondence~\cite{Maldacena:1997re} is used to provide the link between the AdS Randall-Sundrum (RS) models~\cite{Randall:1999ee,Rattazzi:2000hs,Grossman:1999ra,Huber:2000ie,Gherghetta:2000qt,Agashe:2003zs} and strongly coupled scale invariant extensions of the SM.
Holographic techniques have been used in many studies of composite Higgs models to provide a consistent and calculable framework~\cite{Agashe:2004rs,Agashe:2006at,Contino:2006qr} for phenomenology and for predicting relations between the masses and couplings of different states. In the context of the twin Higgs mechanism the holographic realizations are studied in~\cite{Geller:2014kta,Dillon:2018wye,Xu:2019xuo}.
In the holographic picture the field content exists in a 5D spacetime bounded by two D3-branes, and in the bulk there is a negative cosmological constant inducing an AdS curvature which warps 4D energy scales along the extra dimension.
The bulk fields are dual to primary operators of the CFT.
The brane at the IR end of the extra dimension, the IR brane, is associated with the confinement of the new gauge-fermion system of the 4D model. 
While the other brane, the UV brane, is associated with the states external to CFT.
The global symmetries on each brane must respect this correspondence.
The 5D fields in the bulk each give rise to a tower of massive 4D Kaluza-Klein (KK) modes.
It is the presence of the IR brane that induces this mechanism, thus it is associated both with the spontaneous breaking of $\mathcal{G}$ and the spontaneous breaking of the approximate scale invariance.
Fluctuations in the position of the stabilized IR brane are identified with a physical field, the radion~\cite{Goldberger:1999uk}, which through the AdS/CFT correspondence is associated with the dilaton in 4D~\cite{Rattazzi:2000hs}.
Therefore holography presents a consistent and predictive framework in which to study the interplay between the dilaton and composite twin Higgs sectors. 

It is important to emphasize that the holographic models also permit an elegant description of the partial compositeness, with the SM and twin sector fermions having exponentially localized wavefunctions in the bulk entirely controlled by a 5D mass parameter~\cite{Grossman:1999ra,Huber:2000ie}.
The composite states (vector-like KK fermions and KK bosons) and the Higgs field have wavefunctions localized in the IR, therefore through the 5D mass parameters one can control the couplings of the SM fermions to these states.
Note that $\mathcal{O}(1)$ variations in the 5D mass parameters for the SM and twin fermions corresponds to exponential shifts in their couplings to both the Higgs and the composite states.
This allows for natural exponential hierarchies in the flavour sector and for a natural suppression of KK-mediated flavour violating couplings for lighter fermions.

The paper is structured as follows: In the next section we present an effective theory for the SITH model with an approximately scale invariant UV description, where we give a brief overview of the model and introduce the dilaton field as a nonlinear realization of the spontaneously broken scale invariance. We then present the EFT predictions for the dilaton couplings to the light fields within the model and in Sec.~\ref{sec:holo_twin} we implement the holographic realization of the SITH model. 
Finally, in Sec.~\ref{sec:pheno} we present a detailed phenomenological study of the dilaton portal SITH model, where our analysis is based on the LHC Run-2 data for direct searches and indirect constraints on the model parameters.
We present our conclusions in Sec.~\ref{sec:conc}. We supplement the work with an appendix~\ref{sec:appendix}  containing the Feynman rules of three point vertices in the model .

\section{Effective field theory of the SITH}
\label{sec:eff_ccth}

\noindent We consider a SITH model where the UV completion is approximately scale invariant.
To study the low energy phenomenology we use an EFT where heavy composite states have been integrated out.
In the effective Lagrangian we are then left with just the SM states, the twin states, and a light pseudo-Goldstone boson associated with the spontaneous breaking of approximate scale invariance, the dilaton.

\subsection{The SITH model}
\label{sec:CTH}
We assume that the Higgs field arises as a pseudo-Goldstone boson of an approximate $SO(8)$ global symmetry, spontaneously broken to $SO(7)$ at the scale $f$~\cite{Geller:2014kta,Barbieri:2015lqa,Low:2015nqa}.
This global symmetry is broken explicitly by couplings of the SM and twin SM matter fields with the Higgs doublet, which in the end results in a loop-induced potential for the Higgs doublet.
We parameterize this spontaneous breaking by acquiring VEV with the field
\beq
\Sigma=\Sigma_0 e^{i\Pi/f}, \lsp {\rm where}\lsp \Pi=\sqrt2 h^{\hat a} T^{\hat a}, \hsp {\rm and} \hsp \hat a=1,\ldots, 7,
\eeq
where $\Sigma_{0}\!\equiv\!\langle\Sigma\rangle\!=\!f(0,0,0,0,0,0,0,1)^\intercal$, $T^{\hat a}$ are the broken generators of $SO(8)\!/\!SO(7)$ coset, and $h^{\hat a}$ are the corresponding seven Goldstone bosons.
The specific global symmetry for the electroweak sector in this model is
\begin{align}\label{ewsymmetry}
SO(8)\times U(1)_X\times U(1)_{\hat{X}}&\supset SO(4)\times SO(4)\times U(1)_X\times U(1)_{\hat{X}}\nonumber\\ 
&\simeq SU(2)_{L}\times SU(2)_{R}\times SU(2)_{\hat{L}}\times SU(2)_{\hat{R}}\times U(1)_X\times U(1)_{\hat{X}}.
\end{align}
The two $SO(4)$ groups correspond to the custodial groups for the SM Higgs and the twin Higgs doublets, and {\it hatted}\, indices on the subscripts imply that the subgroup is associated with the twin sector.
From this global symmetry we can gauge subgroups corresponding to the spin-1 fields of the SM and twin sectors. We gauge $SU(2)_L\!\times\! U(1)_Y$ and its twin $SU(2)_{\hat L}\!\times\! U(1)_{\hat Y}$, where the SM and twin hypercharges are defined as $Y=T_R^3+\frac43 X$ and  $\hat Y=T_{\hat R}^3+\frac34\hat X$, respectively. 
As $SO(8)$ is spontaneously broken to $SO(7)$, the $SO(4)$ subgroup corresponding to the custodial symmetry of the twin Higgs is broken to $SO(3)$.
This means that there will be a heavy radial mode from the breaking corresponding to the twin Higgs doublet, and a massless Goldstone $SO(4)$ multiplet corresponding to the SM Higgs doublet.
The radial mode of the twin Higgs doublet is integrated out along with other heavy composite states\,\footnote{In weakly coupled twin Higgs scenarios the radial mode can be relatively light. The phenomenology of such state is considered in Refs.~\cite{Ahmed:2017psb,Chacko:2017xpd}.}.
In the unitary gauge, six of the Goldstone bosons are `eaten' by the SM and twin weak gauge bosons ($W^\pm,Z,\hat W^\pm,\hat Z$) and the $\Sigma$ field takes the form
\begin{equation}
\Sigma=f\left(0,0,0,s_h,0,0,0,c_h\right)^\intercal, \lsp {\rm with }~ s_h\equiv\sin(h/f), ~c_h\equiv\cos(h/f),
 \end{equation}
where $h\equiv \sqrt{2 H^\dag H}$ is the Higgs boson, and $H$ is the SM Higgs doublet under $SU(2)_L$.

The model also requires a twin QCD sector with the same gauge coupling as the SM QCD.
This is achieved through embedding the SM QCD and its twin copy in a global $SU(7)$ symmetry \cite{Geller:2014kta}, and then gauging the subgroup $SU(3)_c\!\times\!SU(3)_{\hat c}$.
The SM QCD gauge group is described by $SU(3)_c$ while $SU(3)_{\hat c}$ describes twin QCD.
The SM and twin matter fields are introduced as being external to the strongly coupled sector.
To write down interactions between the SM and twin matter fields we must embed them in same representation of the global symmetries.
Therefore SM and twin matter fields of the same species are embedded in the same multiplet, giving rise to the $Z_2$ symmetry between the two sectors which results in the softening of the Higgs potential described in the introduction.
The embeddings of the SM and twin quarks in the $SO(8)$ global symmetry are $Q_L/\hat{Q}_L\in {\bf 8}$, $t_R/\hat{t}_R\in {\bf 1}$, and $b_R/\hat{b}_R\in {\bf 28}$, with each of them being embedded in a ${\bf 7}$ of $SU(7)$ \cite{Geller:2014kta}.
It is straight-forward to check that under the decomposition of $SO(8)$ in Eq.~\eqref{ewsymmetry} these embeddings result in SM and twin states with the correct quantum numbers. 
There are heavy resonances associated with the composite sector at a scale $m_\ast$ which are charged under both the SM and twin sectors.
We define the scale $m_\ast \equiv g_\ast f$ with $g_\ast$ being the strong coupling of the composite sector which has the generic size $1\lesssim g_\ast\lesssim4\pi$. 

Below scale $m_\ast$ the model is described by the following EFT,
\begin{align}\label{theft}
\mathcal{L}_{\text{\tiny TH}}
&=-\frac{1}{4}\sum_i\left(F_i^{\mu\nu}F_{i\mu\nu}+F_{\hat{i}}^{\mu\nu}F_{\hat{i}\mu\nu}\right)+\sum_\psi\left(i\bar{\psi}\slashed{D}\psi + i\bar{\hat{\psi}}\slashed{D}\hat{\psi} - m_{\psi}\bar{\psi}\psi-m_{\hat{\psi}}\bar{\hat{\psi}}\hat{\psi}\right) \nonumber\\
&\quad +\frac{1}{2}\sum_V\left(m_V^2V^\mu V_{\mu}+m_{\hat{V}}^2\hat V^\mu \hat V_{\mu}\right)+\frac{1}{2}\left(\partial_\mu h\right)^2-V(h^2)\nonumber \\
&\quad +\sum_V\left(vc_{hii}A_i^\mu A_{i\mu}+vc_{h\hat{i}\hat{i}}A_{\hat{i}}^\mu A_{\hat{i}\mu}\right)h -\sum_\psi\left(vy_\psi\bar{\psi}\psi+vy_{\hat{\psi}}\bar{\hat{\psi}}\hat{\psi}\right) h +\ldots,
\end{align}
where $i=(W^{\pm},Z,\gamma,g)$ and $V=(W^{\pm},Z)$ for the SM gauge boson and, similarly, the {\it hatted} states correspond to the twin sector gauge bosons~\footnote{The cosmology of the mirror twin Higgs models introduces some problems which can be addressed either by additional $\mathbb{Z}_2$ breaking~\cite{Chacko:2016hvu,Craig:2016lyx} or by completely decoupling the light states from the twin sector which do not contribute in alleviating the quadratic divergences of the SM Higgs~\cite{Craig:2015pha}.  This issue does not directly influence our results therefore we do not discuss it further.}.  
We have chosen $\psi$ to run over all SM fermions, with $\hat{\psi}$ representing the twin fermions, and we have  integrated out the twin Higgs radial mode. The ellipsis above represent higher order interactions among the light states. 
Note that the masses of the SM and twin states are related as,
\beq
m_{\rm twin}=\frac{\hat v}{v} m_{\text{\tiny SM}}, \lsp {\rm where }\hsp \hat v\equiv \sqrt{f^2-v^2}.
\eeq
Reproducing the masses of the SM electroweak gauge bosons fixes their interactions with the Higgs boson, i.e.
\beq
c_{hVV}=\sqrt{1-s_{\langle h\rangle}^2}g_{hVV}^{\text{\tiny SM}},\lsp c_{h\hat{V}\hat{V}}=-\sqrt{1-s_{\langle h\rangle}^2}g_{hVV}^{\text{\tiny SM}}, \lsp{\rm with} \hsp s_{\langle h\rangle}\equiv\frac{v}{f}\,,
\eeq
where $g_{hVV}^{\text{\tiny SM}}$ is the SM Higgs gauge bosons coupling. 
The scale $f$ is determined by the Higgs potential in Eq. \ref{theft} and is expected to be in the range $\sim\![600\!-\!1200]$ GeV so as to not introduce unacceptable levels of fine-tuning.
Note that in the above Lagrangian we have not included any light vector-like top-partner quarks which are usually present in composite Higgs models.
This is because in the twin Higgs paradigm these light top-partners are not required in order to have a light Higgs without fine-tuning, this is instead achieved through the presence of the twin states, and the $\mathbb{Z}_2$ symmetry between the SM and twin states in the unbroken electroweak phase.
More details on the masses and couplings in the effective Lagrangian will be discussed in Section \ref{sec:holo_twin} where an explicit derivation is discussed.

\subsection{A light dilaton in the SITH}
\label{sec:eff_twin_dilaton}
As mentioned above, we assume approximate scale invariance in the UV strongly coupled model underpinning the twin Higgs mechanism, where at the scale $\Lambda_\cF$ the scale invariance is spontaneously broken.
This scale is determined entirely by the UV dynamics, and is expected to be close to the scale $\Lambda_f\!=\!4\pi f$.
The spontaneous breaking of approximate scale invariance gives rise to a light pseudo-Goldstone boson, the dilaton $\phi(x)$.
Below the scale $\Lambda_\mathcal{F}$ the approximate scale invariance is nonlinearly realized, such that under the scale transformation $x^\mu\to x^{\prime\mu}\!=\!e^{-\lambda} x^\mu $, the dilaton undergoes a shift $\phi(x)\to\phi^\p(x^\p)\!=\!\phi(x)+\lambda {\cal F} $, where $\lambda$ is the scaling parameter. We can then write the dilaton field as a `conformal compensator',
\beq
\chi(x)={\cal F}  e^{\phi(x)/{\cal F} }\,,
\eeq
such that it transforms linearly under the scale transformation, i.e. $\chi(x)\to\chi^\p(x^\p)\!=\!e^{\lambda}\chi(x)$. The VEV of $\chi(x)$, i.e. $\langle \chi(x)\rangle\!\equiv\!{\cal F}$, sets the scale of spontaneous conformal symmetry breaking. 
This is useful in order to derive the low energy interactions between the dilaton and the twin Higgs sector.

Under a scale transformation $x^\mu\to x^{\prime\mu}\!=\!e^{-\lambda} x^\mu$ an operator $\op_i(x)$ with scaling dimension $\Delta_i$ transforms as, 
\beq
\op_i(x)\to\op_i^\prime(x^\prime)= e^{\lambda\Delta_i}\op_i(x).
\eeq
One can then construct a scale invariant effective theory containing operators with $[\Delta]\neq4$ and the appropriate conformal compensator factors $\chi/\cF$.
Without specifying the details of the UV strongly coupled theory we can derive the structure of the interactions in the effective theory through consideration of operator dimensions in the UV and IR, i.e. the UV description and the effective twin Higgs description.
The general form of the approximately scale invariant SITH Lagrangian is
\begin{align}
{\cal L}={\cal L}_{\text{\tiny CFT}}+{\cal L}_{\text{def}}+{\cal L}_{\text{\tiny SM}}^{\rm elem}+\hat {\cal L}_{\text{\tiny TS}}^{\rm elem}+{\cal L}_{\text{mix}},
\end{align}
where ${\cal L}_{\text{\tiny CFT}}$ is the CFT Lagrangian and ${\cal L}_{\text{\tiny SM}}^{\rm elem}(\hat {\cal L}_{\text{\tiny TS}}^{\rm elem})$ is the SM(twin) Lagrangian with elementary fields. Whereas, the ${\cal L}_{\text{def}}$ captures the dynamics that explicitly break scale invariance and has the general form,
\beq
{\cal L}_{\text{def}}=\sum_i g^{\text{\tiny UV}}_i(\mu)\op^{\text{\tiny UV}}_i(x)
\eeq
with $\op^{\text{\tiny UV}}$ being the  primary operators which exhibit explicit breaking of scale invariance in the UV, i.e. $\Delta^{\text{\tiny UV}}_i\neq4$. The Lagrangian ${\cal L}_{\text{mix}}$ includes the mixing of the SM $\op_i^{\text{\tiny SM}}$ and twin sector $\op_i^{\text{\tiny TS}}$  operators with the CFT composite $\op_i^{\text{\tiny CFT}}$ operators and can be written as,
\beq
{\cal L}_{\text{mix}}=\sum_i y_i(\mu)\op^{\text{\tiny SM}}_i(x)\op^{\text{\tiny CFT}}_i(x)+\sum_i \hat y_i(\mu)\hat\op^{\text{\tiny TS}}_i(x)\op^{\text{\tiny CFT}}_i(x)\,.
\eeq
In the following we confine ourselves to a low energy effective theory of this model and for more formal discussion of the CFT dynamics we refer the readers to Refs.~\cite{Chacko:2012sy,Coradeschi:2013gda,Bellazzini:2013fga}.

Assuming approximate scale invariance in the UV a consistent low energy effective theory can be formulated by appropriately incorporating the conformal compensator factors in the effective Lagrangian to ensure the restoration of scale invariance.
In the $SO(8)/SO(7)$ SITH model considered here this restoration occurs through the rescaling $f\to f \,\chi/\!\cF$, such that the nonlinear sigma model condition becomes $|\Sigma|^2=f^2\,\chi^2\!/\!\cF^2$.
Therefore the kinetic term for the nonlinear sigma model would become,
\beq
\frac12 (D_\mu \Sigma)^\dag (D^\mu \Sigma)\to \frac12 (D_\mu \Sigma)^\dag (D^\mu \Sigma)\left(\!\frac{\chi}{\cF}\!\right)^2,	\label{eq:nlsm_chi}
\eeq 
where $D_\mu$ is the covariant derivative in the twin Higgs model. Note that in our conventions $h/\!f$ has zero (classical) scaling dimension, therefore it would not require a conformal compensator factor. After adopting this scheme one can write down an effective theory where all the heavy resonances of the strong dynamics are integrated out and in the low energy theory one recovers a twin Higgs model with a light dilaton $\phi(x)$. 
The explicit breaking of scale invariance in the UV generates a mass for the dilaton, we assume this breaking to be small which translates to the assumption that $m_\phi\!\ll \!\mathcal{F}$.
Note that the explicit breaking of the scale invariance can be naturally small, if the operators responsible for explicit breaking are nearly marginal or their couplings are small, see e.g.~\cite{Goldberger:2008zz,Fan:2008jk,CPR:2010,Chacko:2012sy,Coradeschi:2013gda,Bellazzini:2013fga}. 
In this work we consider the case of near marginal deformations, i.e. operators in the UV with scaling dimensions $\epsilon\!\equiv\!4-\Delta\!\ll\!1$.
 Furthermore, we assume that the beta function for the couplings of these operators do not grow large in the IR, i.e. $\beta(g^{\text{\tiny UV}})\!\sim\!\epsilon$. These two properties ensure a relatively light dilaton whose mass squared is proportional to $\epsilon$, i.e. $m_\phi^2\!\sim\!\epsilon \,\cF^2$~\cite{CPR:2010}.
In this work we assume the mass of the dilaton as a free parameter and require $m_\phi\!\ll\! \cF$. 

To capture the physics associated with the dilaton portal between the SM and twin sector we consider operators up to dimension five. We write the effective SITH model with the dilaton as, 
\begin{align}
{\cal L}= {\cal L}_{\text{\tiny TH}}+{\cal L}_{\rm dilaton}+{\cal L}_{\rm portal}, 	\label{eq:lag}
\end{align}
where ${\cal L}_{\text{\tiny TH}}$ is the Lagrangian containing for the SM and the twin sector given in Eq.~\eqref{theft}. 
The dilaton Lagrangian is
\begin{align}
{\cal L}_{\rm dilaton}&=\frac12 \partial_\mu \chi \partial^\mu \chi-V_{\rm eff}(\chi)
=\frac12 \partial_\mu \phi \partial^\mu \phi-\frac12 m_\phi^2 \phi^2+\cdots,	\label{eq:potdil}
\end{align}
where the ellipses in Eq.~\eqref{eq:potdil} denote higher order dilaton self-interactions, and the effective potential has the form,
\begin{align}
V_{\rm eff}(\chi)&=\frac{\epsilon}{4} \chi^{4}\left[\ln \left(\frac{\chi}{\cF}\right)-\frac{1}{4}\right].	\label{eq:effpotdil}
\end{align}
For detailed discussion on the form of the effective dilaton potential see e.g. Ref.~\cite{Chacko:2012sy}. 
In the following we summarize the dilaton interactions with the SM and twin sector fields, where we neglect terms proportional to $\epsilon=m_\phi^2/\!\cF^2$.

\paragraph{Couplings to massive gauge bosons:} 
The coupling of the dilaton to the massive gauge boson of the SM and twin sectors arise from the gauge covariant kinetic term of nonlinear $\Sigma$ field~\eqref{eq:nlsm_chi},
\[
\frac12 \left(\!\frac{\chi}{\cF}\!\right)^2(D_\mu \Sigma)^\dag (D^\mu \Sigma).
\] 
In the unitary gauge and working to linear order in the dilaton field, we get
\begin{align}
{\cal L}_{\rm portal}\supset \frac{\phi}{{\cal F} }&\bigg[2m_{W}^2 W^{+}_\mu W^{-\mu} + m_{Z}^2 Z_\mu Z^\mu 	
	+2m_{\hat W}^2 \hat W^{+}_\mu \hat W^{-\mu} +m_{\hat Z}^2 \hat Z_\mu \hat Z^\mu\bigg].
\end{align}
Here the mass terms capture the explicit breaking of scale invariance. 

\paragraph{Couplings to massless gauge bosons:} 
The dilaton couplings to massless gauge bosons in both sectors are associated with the breaking of scale invariance due to the running of the gauge couplings and are given by,
\begin{align}
{\cal L}_{\rm portal}\supset \frac{\phi}{{\cal F} }&\bigg[\frac{\alpha_\textsc{qcd}}{8\pi}b_g^{\rm eff} G_{\mu\nu}^a G^{a\mu\nu}+\frac{\alpha_\textsc{em}}{8\pi}b_{\gamma}^{\rm eff} F_{\mu\nu}F^{\mu\nu}+\frac{\alpha_\textsc{em}}{4\pi}b_{Z\gamma}^{\rm eff} F_{\mu\nu}Z^{\mu\nu} 		\notag\\
	&+\frac{\hat \alpha_\textsc{qcd}}{8\pi}\hat b_{\hat g}^{\rm eff} \hat G_{\mu\nu}^a \hat G^{a\mu\nu}+\frac{\hat \alpha_\textsc{em}}{8\pi}\hat b_{\hat \gamma}^{\rm eff} \hat F_{\mu\nu}\hat F^{\mu\nu}+\frac{\hat \alpha_\textsc{em}}{4\pi}\hat b_{\hat Z\hat \gamma}^{\rm eff}\hat F_{\mu\nu}\hat Z^{\mu\nu}\bigg],
\end{align}
where $b_i$ terms are the beta function coefficients, defined as $\beta_i(g_i)\!=\!b_i g_i^3/(16\pi^2)$.
In this model they are given as
\beq
\begin{aligned}
b_g^{\rm eff} &=  b_3^{\text{\tiny<}}- b_3^\text{\tiny>}	,	&\hat b_{\hat g}^{\rm eff} &= \hat b_{3}^\text{\tiny<}- \hat b_3^\text{\tiny>}	,	\\
b_{\gamma}^{\rm eff} &=b_\textsc{em}^\text{\tiny<}- b_\textsc{em}^\text{\tiny>}	,	&\hat b_{\hat \gamma}^{\rm eff} &=\hat b_\textsc{em}^\text{\tiny<}- \hat b_\textsc{em}^\text{\tiny>},	\\
b_{Z\gamma}^{\rm eff} &= (b_{2}^\text{\tiny<}-b_2^\text{\tiny>})\!/\!t_{\theta}- (b_{1}^\text{\tiny<}-b_1^\text{\tiny>})t_{\theta}	,	\lsp	&\hat b_{\hat Z\hat \gamma}^{\rm eff} &=(\hat b_{2}^\text{\tiny<}-\hat b_2^\text{\tiny>})\!/\!t_{\hat \theta}-(\hat b_{1}^\text{\tiny<}-\hat b_1^\text{\tiny>})t_{\hat \theta},
\end{aligned}	\label{eq:bfun}
\eeq
where $b_3, b_2, b_1$ are the beta-function coefficients for the $SU(3)_c,SU(2)_L,U(1)_Y$ gauge groups, respectively. 
We have also defined $b_\textsc{em}\!\equiv\!b_1+b_2$ and $t_\theta\!\equiv\!\tan\theta_W$, with $\theta_W$ being the Weinberg angle. The $b_i^\text{\tiny<}$ and $b_i^\text{\tiny>}$ above correspond to the IR and UV contributions to the beta-function coefficients for the energy scales $0\leq\mu\leq\Lambda_\ir$ and $\Lambda_\ir\leq\mu\leq\Lambda_\uv$, respectively. The specific values of the $b_i$ coefficients are model dependent, and in the next section we will present a UV complete holographic model in which these coefficients can be calculated. 

\paragraph{Couplings to partially composite fermions:} 
Partial compositeness requires that  the elementary fermions $Q, U$ mix with the states associated with the strong dynamics operators ${\cal Q},{\cal U}$, as
\beq
{\cal Y}_{\cal Q}^{i\alpha}Q_i {\cal Q}_\alpha^c+{\cal Y}_{\cal U}^{\beta j} {\cal U}_\beta U_j^c+\hc,
\eeq
where $\alpha,\beta=1,2,3$ and ${\cal Y}_{\cal Q}$ are the mixing matrices. The strong sector fermionic operators ${\cal Q},{\cal U}$ transform as {\bf 8} and {\bf 1} under the $SO(8)$ symmetry with scaling dimensions $\Delta_{\cal Q}$ and $\Delta_{\cal U}$, respectively. At the linear order in dilaton field, the dilaton couplings to partially composite states are given by,
\begin{align}
{\cal L}_{\rm portal}\supset -\frac{\phi}{f} \Big[m^{i j} [\Delta_{\cal Q}+\Delta_{\cal U}-4]q_{i} u_{j}^{c}+ \hat m^{i j} [\Delta_{\cal Q}+\Delta_{\cal U}-4]\hat q_{i} \hat u_{j}^{c},
\end{align}
where $m^{i j}$ and $\hat m^{i j}$ are mass matrices for the light SM and twin sector fermions, respectively. 

\paragraph{Couplings to SM-like Higgs boson:} 
The dilaton couplings to the SM-like  Higgs  boson, which is a fully composite state, can be obtained from the kinetic term of the nonlinear $\Sigma$ field at the linear order in the dilaton field,
\begin{align}
{\cal L}_{\rm portal}\supset \frac{\phi}{f} \Big[\partial_{\mu} h\partial^{\mu} h+ 2m_h^2\,h^2\Big],
\end{align}
where the explicit scale invariance breaking effects to the kinetic and mass terms of the Higgs are assumed to be small compared to the leading terms\,\footnote{The explicit scale invariance breaking effects for the psuedo-Goldstone Higgs depend on the SITH model and, in particular, the details of the $\mathbb{Z}_2$ soft breaking terms in the model. For phenomenological purposes, in Sec.~\ref{sec:holo_twin} we employ a holographic twin Higgs model~\cite{Geller:2014kta}, where such $\mathbb{Z}_2$ breaking terms are introduced holographically and the parametric running effects on these terms are assumed to be small. For our purposes here we assume these terms do not introduce large explicit scale invariance breaking effects to the SM Higgs potential.} and hence neglected. The dilaton-Higgs kinetic and/or mass mixing terms are model dependent~\cite{Chacko:2012sy} and in the models considered here they are small enough to be neglected, as we will discuss in the next section.

\section{A holographic SITH model}
\label{sec:holo_twin}
In this section we present an explicit example where the dilaton portal is realized in the twin Higgs model and we calculate the interactions of the dilaton to the SM and twin sector states.
The model we consider is the 5D holographic UV completion of the twin Higgs mechanism~\cite{Geller:2014kta}, with the radion field taking the role of the dilaton.
The effective theory is derived using holographic methods and will involve the same global symmetry structure presented in Sec.~\ref{sec:eff_ccth}.
Therefore we expect the effective theory to have the same structure, however with the advantages of calculability provided by the holographic model.

\begin{figure}[t!]
\centering
\begin{tabular}{cl}
\raisebox{-.5\height}{\includegraphics[width=0.37\textwidth]{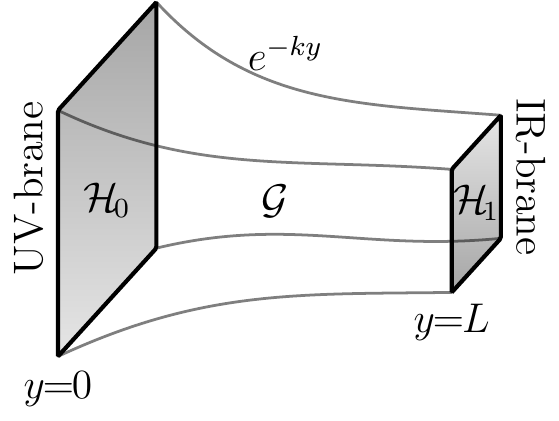}}& \parbox{3cm}{\begin{align*}{\cal G}\!&=\!SU(7)\!\times\! SO(8)\!\times\!\Z2 \\
{\cal H}_1\!&=\!SU(7)\!\times \!SO(7)\!\times\! \Z2 \\
{\cal H}_0\!&=\!{\cal G}_{\text{\tiny SM}}\!\times\! {\cal G}_{\rm twin}\!\times\! \mathbb{Z}_2\\
{\cal G}_{\text{\tiny SM}}\!&=\!SU(3)_c\!\times\! SU(2)_L\!\times\! U(1)_Y\\
{\cal G}_{\rm twin}\!&=\!SU(3)_{\hat c}\!\times\! SU(2)_{\hat L}\!\times\! U(1)_{\hat Y}
\end{align*}}
\end{tabular}
\caption{A pictorial depiction of holographic twin Higgs which employs five dimensional RS geometry with a bulk gauge symmetry ${\cal G}$ which is broken on the UV-brane to ${\cal H}_0$ and on the IR brane to ${\cal H}_1$.}
\label{fig:rs}
\end{figure}

The holographic twin Higgs model~\cite{Geller:2014kta} consists of an RS geometry (see Fig.~\ref{fig:rs}) with a bulk gauge symmetry, ${\cal G}\!=\!SU(7)\!\times\! SO(8)\!\times\!\Z2$,  which is broken on the IR brane to ${\cal H}_1\!=\!SU(7)\!\times \!SO(7)\!\times\! \Z2$ and on the UV brane to ${\cal H}_0\!=\![SU(3)_c\!\times\! SU(2)_L\!\times\! U(1)_Y]\!\times\! [SU(3)_{\hat c}\!\times\! SU(2)_{\hat L}\!\times\! U(1)_{\hat Y}]\!\times\! \mathbb{Z}_2$.
The QCD gauge group and its twin arise through $SU(7)\supset SU(3)_{c}\times SU(3)_{\hat{c}}\times U(1)_{X}\times U(1)_{\hat{X}}$, with the {\it hats} representing the twin groups.
The electroweak gauge group and its twin then arise through $SO(8)\supset SO(4)\times SO(4)\simeq SU(2)_{L}\times SU(2)_{R}\times SU(2)_{\hat{L}}\times SU(2)_{\hat{R}}$, with SM hypercharge and twin hypercharge being defined as $Y=T^3_R+\frac{4}{3}X$ and $\hat{Y}=T^3_{\hat{R}}+\frac{4}{3}{\hat{X}}$.
Because the SM and twin fields arise from the same bulk gauge group, their bulk gauge couplings are identical.
The breaking of $SO(8)$ to $SO(7)$ on the IR brane gives rise to seven Goldstone bosons, four of these transform as a bi-doublet of the SM sector $SU(2)_L\!\times\! SU(2)_R$ and assume the role of the SM Higgs doublet, while the other three are eaten (in the unitary gauge) by the twin gauge bosons ($\hat W^\pm, \hat Z$) of the gauged $SU(2)_{\hat{L}}\!\times\!U(1)_{\hat Y}$. 
The twin Higgs radial mode, which spontaneously breaks $SO(8)$ to $SO(7)$ at the scale $f$, is integrated-out of the low-energy effective theory by using the boundary conditions.
The bulk global symmetry is broken explicitly on the UV brane through the gauging of SM and twin sector gauge symmetries, which in turn generates a non-trivial SM Higgs potential. It was shown that a realistic implementation of EWSB can be achieved with $m_\ast\!=\!\mkk\!\sim\!\mathcal{O}(5)$ TeV.  The measured SM-like Higgs boson mass can be reproduced in the model with the inclusion of a $\mathbb{Z}_2$ breaking term \cite{Geller:2014kta}, we will therefore keep the Higgs mass fixed and focus on the unexplored phenomenology of the dilaton portal in this SITH model. 

The holographic model employs a 5D RS geometry which enforces an exponential hierarchy in scales through an extra dimension with an Anti-de Sitter (AdS) background. The extra dimension is bounded by two D3-branes whose tensions ensure 4D Poincare invariance at every point in the bulk, i.e. the space between the branes, see Fig.~\ref{fig:rs}.
The metric describing this warped extra dimension is
\beq\label{metric2}
ds^2=e^{-2ky}\eta_{\mu\nu}dx^\mu dx^\nu-dy^2,
\eeq
with $k$ is the curvature scale and $y$ being the extra dimensional coordinate. The branes cut off the extra dimension at $y=0$ and $y=L$, and we refer to these as the UV- and IR-branes, respectively. Fields can live either on the branes or in the whole 5D bulk, and may have different localizations determined by the Lagrangian describing their dynamics.
The presence of the branes means that each field in the bulk can be decomposed through a Kaluza-Klein (KK) decomposition into discrete 4D mass eigenstates.
Aside from a possible massless zero mode, the lightest of these modes will be near the KK scale, $\mkk\!\equiv\!2k\,e^{-kL}$, with the exact mass spectrum depending on the field's spin and the 5D mass term.
The AdS/CFT correspondence relates the RS model to a strongly coupled gauge theory in 4D, whose conformal invariance is spontaneously broken near the KK scale, which in turn is dictated by the position of the IR brane.
In building a holographic/composite Higgs model which renders the Higgs sector natural we expect the position of the IR brane to be at $L\approx 35/k$ (if $k\sim \mpl$) such that the KK scale is close to~${\cal O}({\rm TeV})$.

Without a mechanism to dynamically fix the position of the IR brane a massless state, called the radion, exists in the spectrum and corresponds to fluctuations of the inter-brane distance $L$. In the dual picture, via AdS/CFT duality, this state corresponds to a Goldstone boson associated to spontaneous breaking of the scale/conformal invariance of the 4D CFT theory --- the dilaton~\cite{Rattazzi:2000hs,Chacko:2014pqa}.  We employ the well-known Goldberger-Wise stabilization mechanism~\cite{Goldberger:1999uk} to dynamically fix the position of the IR brane, which in turn gives rise to a mass for the dilaton/radion state.
The Goldberger-Wise mechanism employs a bulk scalar field whose potential is minimized for some finite value of the extra dimensional length~$L$.
The Einstein-Hilbert action for such a set-up can be written as
\beq\label{EH}
S=\int dx^4\int_0^Ldy~\sqrt{-g}\left[2M_*^3 R-\frac{1}{2}g^{MN}\partial_M\Phi\partial_N\Phi-V(\Phi)\right],
\eeq
where $g_{MN}$ is the 5D metric, $g$ is its determinant, $M_*$ is the 5D Planck mass, $R$ is the Ricci scalar, and $\Phi$ is the Goldberger-Wise field.

Fluctuations around the background metric give rise to spin-0, spin-1, and spin-2 degrees of freedom.
However the presence of the branes induces a spontaneous symmetry breaking mechanism that effectively results in the massive spin-2 modes eating the spin-1 and spin-0 degrees of freedom.
In the end we are left with a massless spin-0 field ({\it the radion}), a massless spin-2 field ({\it the graviton}), and a tower of massive spin-2 fields ({\it the KK gravitons}) with masses of order $\mkk$.
We describe the perturbed metric by $g_{MN}=\bar{g}_{MN}+h_{MN}$ where $\bar{g}_{MN}$ is that described by Eq.~\eqref{metric2}.
To study the fluctuations one starts from the Einstein equation with the energy-momentum tensor for the SM and twin fields, the Goldberger-Wise field, and any other additional field content.
A perturbation series in $h_{MN}$ is then performed and the couplings linear in the radion and KK graviton are extracted.
Since we are only interested in the radion fluctuation we can parameterize the perturbed metric with
\beq
ds^2=e^{-2ky}(1-2\hat{\phi})\eta_{\mu\nu}dx^\mu dx^\nu -(1+4\hat{\phi})dy^2,
\eeq
where $\hat{\phi}(x,y)$ is the radion 5D field.
The physical radion is a mixture of the metric fluctuation and the Goldberger-Wise field $\Phi(x,y)$, however the component of the Goldberger-Wise field is typically negligible and we will assume that limit here.

Decomposing the radion field into its 4D and 5D components as $\hat{\phi}(x,y)=\phi(x)f_\phi(y)$ it is found that, in the zero back-reaction limit, the equation of motion for 5D profile of the radion is
\beq
\partial_5^2f_\phi(y)-2k\partial_5f_\phi(y)=0,
\eeq
and its solution is $f_\phi(y)\!=\!e^{2ky}/{\cal F} $, where ${\cal F} $ is a constant set by the VEV of the radion/dilaton field. We will not provide a thorough review of the Goldberger-Wise mechanism here, and we will simply assume for phenomenological reasons, that the mass of the radion is a free parameter and that the backreaction induced onto the RS metric from the stabilization mechanism is negligible\,\footnote{The radion phenomenology has been studied extensively without twin sector scenarios, see e.g.~\cite{Csaki:2000zn,Dominici:2002jv,Csaki:2007ns,Cox:2013rva,Ahmed:2015uqt}.}. 
This is justified through our assumption that the radion/dilaton mass is much smaller than the KK scale, i.e. $m_\phi\!\ll\!\mkk$, in accordance with the assumptions of the previous section if $\mkk\sim \mathcal{F}$.

The mass of the radion induces a correction to this 5D profile $f_\phi(y)$ of the order $m_\phi^2/\mkk^2$ at the very most and therefore we will neglect it in the following analysis. The kinetic term of the radion is normalized with ${\cal F} ^{2}=12M^3_*\int_0^Ldy~e^{-2ky}$, hence
\beq
{\cal F} \simeq\sqrt{6}\mpl \,e^{-kL}=\sqrt{\frac32}\frac{\mpl}{k}\,\mkk\,,	\label{eq:f_mkk}	
\eeq
where the 4D Planck mass $\mpl$ is related to the 5D Planck mass $M_\ast$ and the curvature scale $k$ as $\mpl^2\simeq M_*^3/k$.
Note that the mass dimension $[\hat{\phi}(x,y)]\!=\!0$, however $[{\cal F} ]\!=\!1$ and $[\phi(x)]\!=\!1$.
Once the extra dimension has been integrated out it is generally found that the radion couplings to massive particles scale as $m_X/{\cal F} $ where $m_X$ is the mass of the particle $X$ in question.
This limit is exact when the SM fields all reside on the IR brane, however here we allow for the fields to propagate into the bulk, for a more complete treatment on the coupling of the radion to bulk fields, see e.g.~\cite{Chacko:2014pqa}.
Despite the radion being neutral it also couples to massless gauge fields via a coupling to the field strength tensor, this arises both through loops of charged matter fields and through the so-called scale anomaly.

An important feature in RS models of EWSB is usually the inclusion of a Higgs-radion mixing term on the IR brane, while such a term on the UV brane is allowed but would be exponentially suppressed by the wavefunction overlaps.
In the case of a pseudo-Goldstone Higgs, as in the twin Higgs models, this term can only be induced at the loop level since the boundary conditions on the IR brane are invariant under the global symmetry.
This invariance implies that any tree-level coupling between the Ricci scalar (from which the radion arises) and the pseudo-Goldstone Higgs is forbidden.
The size of such a term was estimated in~\cite{Chacko:2012sy,Cox:2013rva} to be $\sim\! m_h^2/f ^2$ at one-loop, and assuming current bounds of $f/v \!\gtrsim\! 3$~\cite{Ahmed:2017psb,Chacko:2017xpd} we determine that in a pseudo-Goldstone Higgs scenario this term is negligible and hence we neglect such mixing in this work.

\begin{figure}[t!]
\centering
\includegraphics[width=0.37\textwidth]{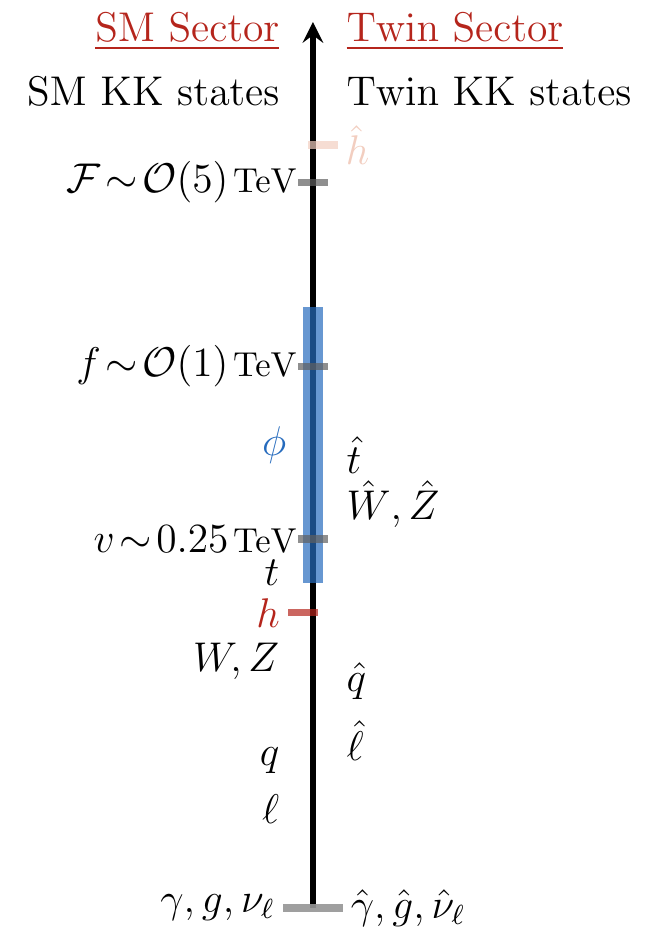}
\caption{Spectrum of the dilaton portal holographic mirror twin Higgs model.}
\label{fig:spectrum}
\end{figure}

Neglecting the radion field for now, we write down the effective Lagrangians for the interactions between the Higgs, top sector, and gauge fields in the holographic SITH model:
\beq
\begin{aligned}
\mathcal{L}_{g}&=\frac{1}{2}P_T^{\mu\nu}\left[W_\mu^+\left(\Pi^W_0+\frac{s_h^2}{2}\Pi^W_1\right)W^{\mu-}+Z_\mu\left(\Pi^W_0+\frac{s_h^2}{2c_W^2}\Pi^W_1\right)Z^\mu+A_\mu\Pi_0^AA^\mu+G^a_\mu\Pi_0^GG^{\mu a}\right], 		\\
\mathcal{L}_{\hat{g}}&=\frac{1}{2}P_T^{\mu\nu}\left[\hat{W}_\mu^+\left(\Pi_0^W+\frac{c_h^2}{2}\Pi_1^W\right)\hat{W}^{\mu-}+\hat{Z}_\mu\left(\Pi_0^W+\frac{c_h^2}{2c_W^2}\Pi_1^W\right)\hat{Z}^\mu+\hat{A}_\mu\Pi_0^A\hat{A}^\mu+\hat{G}^a_\mu\Pi_0^G\hat{G}^{\mu a}\right],		\\
\mathcal{L}_{f}&=\bar{t}_L\cancel{p}\left(\Pi_0^q+\frac{s_h^2}{2}\Pi_1^q\right)t_L + \bar{t}_R\cancel{p}\Pi_0^tt_R-\frac{M_1^ts_h}{\sqrt{2}}(\bar{t}_Lt_R+t_L\bar{t}_R)	,		\\
\mathcal{L}_{\hat{f}}&=\bar{\hat{t}}_L\cancel{p}\left(\Pi_0^q+\frac{c_h^2}{2}\Pi_1^q\right)\hat{t}_L + \bar{\hat{t}}_R\cancel{p}\Pi_0^t\hat{t}_R-\frac{M_1^{\hat{t}}c_h}{\sqrt{2}}(\bar{\hat{t}}_L\hat{t}_R+\hat{t}_L\bar{\hat{t}}_R),	
\end{aligned}
\eeq
where $c_W$ is the cosine of the Weinberg angle, and $s_h\!\equiv\!\sin(h/f )$ and $c_h\!\equiv\!\cos(h/f )$ parameterize the Higgs boson interactions when the Higgs field is nonlinearly realized.
Note that we have assumed the same embeddings of SM and twin fermions in the global symmetry as indicated in section \ref{sec:eff_ccth}.
The form factors $\Pi_{0,1}^{W,A,G,q,t}$ and $M_1^{t,\hat{t}}$ are calculated using 5D holographic techniques and can be found in \cite{Croon:2015wba,Dillon:2018wye}.
The form factors contain poles at the masses of the heavy composite/KK states, however in studying the phenomenology below the TeV scale we need only to look at the low energy behavior of these form factors to extract the couplings to the Higgs boson and the radion.
At low momenta the gauge field form factors can be approximated as
\beq
\begin{aligned}
\Pi_0^W&=\hat{\Pi}_0^W=\frac{p_E^2}{g_5^2k} kL, \lsp 
&\Pi_1^W&=\hat{\Pi}_1^W=-\frac{\mkk^2}{2g_5^2k},		\\
\Pi_0^A&=\hat{\Pi}_0^A=\frac{p_E^2}{g_5^2ks_W^2} kL, 	\lsp
&\Pi_0^G&=\hat{\Pi}_0^G=\frac{p_E^2}{g_5^{c 2}k} kL	,
\end{aligned}
\eeq
where we use the Euclidean momenta $p_E$.
The 5D gauge couplings are then related to the 4D gauge couplings via $g\!=\!g_5/\sqrt{L}$ for the electroweak gauge coupling, and similarly for the others.
Imposing the nonlinear sigma model form factor relation $\Pi_1(p_E\!=\!0)\!=\!-f ^2/2$, we get an expression for the KK resonance masses $\mkk$ in terms of the decay constant $f$ of the twin Higgs field and the effective strong coupling, 
\beq
\mkk = g_\ast f\,, \lsp {\rm where}\lsp g_\ast\equiv g\, \sqrt{kL}.	\label{eq:mkk}
\eeq
In order to have the scale of the IR brane $\op({\rm TeV})$ one requires $kL\approx 35$ which already sets $g_\ast \simeq 6 g$ with $g$ being the SM gauge coupling. However $g_\ast$ can be treated as a free parameter to fix the scale of $\mkk$ for a given value of $f$. 
This allows us to write the dilaton/radion VEV ${\cal F} $ in terms of the scale  $f $ and the 5D parameters through the use of Eq.~\eqref{eq:f_mkk},
\beq
{\cal F} =\sqrt{\frac32}\frac{\mpl}{k}\,g_\ast\,f \,.	\label{eq:cF}
\eeq
Note in deriving the above relations we have made use of the fact that the SM and twin gauge and Yakawa couplings are exactly equal. Fixing the masses of the electroweak gauge bosons to their measured values we see that the VEV of the pseudo-Goldstone Higgs field is related to the SM Higgs VEV and the decay constant through 
\beq
s_{\langle h\rangle}=\frac{v}{f }\,.
\eeq
The mass spectrum of the KK gauge fields are encoded in the poles of the form factors, however these are typically above the TeV scale with the exception of the lowest laying (zero-) modes of the SM and twin gauge fields.
The masses of the lightest twin electroweak bosons are
\begin{align}
m_{\hat{W}}^2=\frac{g^2}{4}(f ^2-v^2)=\frac{\hat v^2}{v^2}m_W^2,	\lsp  m_{\hat{Z}}^2=\frac{g^2}{4c^2_W}(f ^2-v^2)=\frac{\hat v^2}{v^2}m_Z^2\,.
\end{align}
The low energy form factors for the quarks are more complicated due to the 5D mass parameters, however the lightest (zero-mode) twin fermion masses can be expressed straightforwardly as
\begin{align}
m_{\hat{f}}&=\frac{\hat v}{v}m_f\,.
\end{align}
In Fig.~\ref{fig:spectrum}, we sketch the spectrum of holographic twin Higgs model in the presence of a relatively light dilaton/radion state $\phi(x)$ which provides a portal between the SM and twin sectors, along with the usual Higgs portal. 
%

\subsection*{The Higgs coupling to gluons}

The effective coupling of the Higgs to gluons is crucially important for calculating the Higgs production cross-section.
It is well known that in composite Higgs models this effective coupling is modified due to the pseudo-Goldstone nature of the Higgs, and that the correction depends strongly on the form of the couplings to the colored top-partners.
This has been studied both in the 4D \cite{Azatov:2011qy,Carena:2014ria,Banfi:2019xai} and 5D \cite{Falkowski:2007hz} contexts where the results match once one properly considers the presence of all relevant top-partners.
Here we will consider the effective description of our holographic top sector, in which the left-handed top quark resides in an ${\bf 8}$ of $SO(8)$, the right-handed top quark resides in a singlet, and vector-like top-partners appear both in the ${\bf 8}$ and in the singlet.
It suffices here to only consider the $SO(4)$ subgroup of $SO(8)$ corresponding to the SM custodial symmetry, thus we write the SM left-handed doublet embedding as
\beq
Q_L=\frac{1}{\sqrt{2}}
\begin{pmatrix}
ib_L\\b_L\\it_L\\-t_L
\end{pmatrix},
\eeq
with the vector-like top-partners embedded in the fourplet as
\beq
\Psi_4=\frac{1}{\sqrt{2}}
\begin{pmatrix}
iB-\tilde{X}\\B+\tilde{X}\\iT+iX\\-T+X
\end{pmatrix}.
\eeq
The right-handed top, $t_R$, and the remaining vector-like top-partner, $\Psi_1$, are simply singlets of this $SO(4)$.
Denoting the $SO(4)$ Higgs components of the Goldstone multiplet as $\Sigma_4=f(0,0,0,s_h)^T$ in unitary gauge, we can write down the effective theory for the top-sector as
\beq
\mathcal{L}=y\bar{Q}_L\Sigma_4 t_R+y_1\bar{Q}_L\Sigma_4 P_R\Psi_1 +y_2\bar{\Psi}_4\Sigma_4t_R+m_4\bar{\Psi}_4\Psi_4+m_1\bar{\Psi}_1\Psi_1
\eeq
where $y$ and $y_{1,2}$ are all $\mathcal{O}(1)$ couplings generated from the 5D localizations of the left and right-handed top quarks in 5D, and $m_{4,1}$ are the vector-like masses of the lightest top-partner KK states.
The modifications to the effective coupling of the Higgs to gluons then scales with $\tfrac{\partial}{\partial h}\text{log}\left(\text{det} M\right)$ where $M$ is the mass matrix for the top-sector.
To describe these effects we define $R_g=v\tfrac{\partial}{\partial v}\text{log}\left(\text{det} M\right)$ as the ratio of the the Higgs coupling to gluons in the twin Higgs model to the same coupling in the SM.
Assuming $y_1\simeq y_2\simeq y$ and $m_1\simeq m_4\equiv \mkk$ we calculate this to be
\beq
R_g=v\frac{\partial}{\partial v}\text{log}\left(\text{det} M\right)\simeq\sqrt{1-\frac{v^2}{f^2}}\bigg(\frac{1+2y\frac{v}{\mkk}}{1+y\frac{v}{\mkk}}\bigg)\simeq \sqrt{1-\frac{v^2}{f^2}}\bigg(1+\mathcal{O}\Big(y\frac{v}{\mkk}\Big)\bigg).
\eeq
Given that we assume $\mkk\gtrsim \op(5)$ TeV due to the twin Higgs mechanism alleviating fine-tuning constraints, the $v/\mkk$ effects can be safely neglected.
The Higgs coupling to twin gluons can be obtained through a simple exchange of $s_h\leftrightarrow c_h$ due to the $\mathbb{Z}_2$ symmetry,
\beq
R_{\hat{g}}=v\frac{\partial}{\partial v}\text{log}\left(\text{det} \hat{M}\right)\simeq\frac{v}{f}\left(1+\mathcal{O}\left(y\frac{v}{\mkk}\right)\right),
\eeq
where $R_{\hat{g}}$ is the ratio of the SM Higgs coupling with the twin gluons to its coupling with the SM gluons, and $\hat{M}$ is the mass matrix of the twin-top sector.
This is at least valid when only the twin top quark is heavier than the Higgs, which is certainly true for the ranges of $f\!/\!v$ considered in this paper.

\section{Dilaton portal phenomenology}
\label{sec:pheno}

In this section we present a detailed study of the phenomenological implications of the presence of a dilaton/radion portal in the twin Higgs model. 
Although the pseudo-Goldstone Higgs field in the holographic twin Higgs model arises from a bulk gauge field, the dilaton/radion couplings to mass terms are more closely related to the case of a brane-localized Higgs field than a bulk Higgs field. This is because the electroweak symmetry breaking is triggered by boundary conditions on the IR brane, and in fact the pseudo-Goldstone Higgs field can be shifted to the brane via bulk gauge transformations. 

The couplings of the dilaton/radion to the SM and its mirror sector states are straightforward to calculate as outlined in Sec.~\ref{sec:eff_twin_dilaton}.
The  loop-induced and anomalous dilaton couplings to the massless gauge bosons are very important for the dilaton phenomenology.
We collect the most important Feynman rules for the model in Appendix~\ref{sec:appendix}, where the universal Higgs and dilaton couplings $g_h (\hat g_h)$ and $g_\phi(\hat g_\phi)$ to the SM (twin) sector are defined as, 
\begin{align}
g_h\equiv \frac{\hat v}{f }\,, \lsp \hat g_h\equiv -\frac{v}{f }\,,\lsp
g_\phi\equiv \frac{v}{{\cal F} }\,, \lsp \hat g_\phi\equiv \frac{\hat v}{{\cal F} }\,.	\label{eq:gh_gphi}
\end{align}
The phenomenology of the twin Higgs model with a dilaton portal is determined by relatively small set of parameters; $m_\phi, f, \cF$, and $\mkk$; where $f$ sets the  masses of light twin sector states and reduced couplings for the SM Higgs, $\cF$ fixes the interaction strength of the dilaton with both sectors, and $\mkk$ is the overall scale of KK modes of the both sectors which we set to $\op(5)\tev$ such that they are out of the LHC direct reach.  In addition to the above four parameters, other important parameters of the model are the anomalous gauge coupling coefficients $b$'s, which we will discuss in the following.

\subsubsection*{Decay widths and branching ratios}
\label{sec:branching}
Now we set to calculate the partial decay widths and branching ratios of the dilaton/radion to the SM and twin sectors. The two body decays of dilaton into the SM and twin sector fermions are given as:
\beq
\Gamma^{\phi }_{\psi \psi}=\frac{G_{\textsc f}g_{\phi }^2}{4\sqrt2\pi}\mphi  m_\psi^2\bigg(1-\frac{4m_\psi^2}{\mphi ^2}\bigg)^{3/2}, \lsp
\Gamma^{\phi }_{\hat \psi \hat  \psi}=\frac{\hat G_{\textsc f}\hat g_{\phi }^2}{4\sqrt2\pi}\mphi  m_{\hat  \psi}^2\bigg(1-\frac{4m_{\hat  \psi}^2}{\mphi ^2}\bigg)^{3/2},   \label{drate_hff}
\eeq
where $g_{\phi }\!=\!v/{\cal F} $, $\hat g_{\phi }\!=\!\hat v/{\cal F} $, and the twin Fermi constant is $\hat G_{\textsc f}\!\equiv \!(v^2/\hat v^2)G_{\textsc f}$. 
In the following numerical analysis for simplicity, we take $\Delta_{\cal Q}=\Delta_{\cal U}=5/2$ for the SM and twin sector fermions.  
This choice of scaling dimensions makes the dilaton couplings proportional to the fermion masses. 
In general, for light fermions these scaling dimensions differ from $5/2$, however, such deviations are less relevant for the dilaton phenomenology.   
Moreover, in our numerical analysis we have allowed off-shell decays of the heavy fermions (the SM and the twin top quarks).  

The dilaton/radion partial decay width to on-shell massive SM gauge bosons is, 
\begin{align}
\Gamma^{\phi }_{V_1V_2}&=\frac{G_{\textsc f}\,\mphi ^3\, g_{\phi }^2}{8\sqrt2\,\pi\,{\cal S}_V} \lambda\big(x_1,x_2\big) \Big[\lambda^2\big(x_1,x_1\big)
    +12 \,x_1\,x_2\Big] ,
\end{align}
where $\lambda(x_1,x_2)\equiv\sqrt{(1-x_1-x_2)^2-4x_1 x_2}$ with $x_{1,2}=m_{V_{1,2}}^2/\mphi ^2$. ${\cal S}_V$ is a symmetry factor which for identical gauge bosons is 2, while  otherwise it is 1. 
We also include the dilaton decays with off-shell massive gauge bosons, which allows for three and four-body decays of the dilaton. Similarly, for the twin massive gauge bosons partial withs  $\Gamma^{\phi }_{\hat V_1\hat V_2}$ are given by the analogous expressions by replacing $G_{\textsc f}\to \hat G_{\textsc f}$, $g_{\phi}\to \hat g_{\phi}$,  and $m_{V}\to m_{\hat V}$.  
The partial decay widths of the dilaton to the SM and twin sector massless gauge bosons are,
\begin{align}
\Gamma^{\phi }_{gg}&=\frac{G_{\textsc f}\alpha_{\textsc{qcd}}^2\mphi ^3}{16\sqrt2\pi^3}\,c_{g}^2,	
&\Gamma^{\phi }_{\hat g\hat g}&=\frac{\hat G_{\textsc f}\hat \alpha_{\textsc{qcd}}^2\mphi ^3}{16\sqrt2\pi^3}\,\hat c_{\hat g}^2,	\\
\Gamma^{\phi }_{\gamma\gamma}&=\frac{G_{\textsc f}\alpha_{\textsc{em}}^2\mphi ^3}{128\sqrt2\pi^3}\, c_\gamma^2,	
&\Gamma^{\phi }_{\hat \gamma\hat \gamma}&=\frac{\hat G_{\textsc f}\hat \alpha_{\textsc{em}}^2\mphi ^3}{128\sqrt2\pi^3}\,\hat c_{\hat\gamma}^2,\\
\Gamma^{\phi }_{Z\gamma}&=\frac{G_{\textsc f}^2\alpha_{\textsc{em}}\mphi^3  m_Z^2}{16\pi^4}s_{\theta}^2c_{\theta}^2\Big(1-\tfrac{m_Z^2}{\mphi ^2}\Big)^3\; c_{Z\gamma}^2,  
&\Gamma^{\phi }_{\hat Z\hat \gamma}&=\frac{\hat G_{\textsc f}^2\hat \alpha_{\textsc{em}}\mphi^3  m_{\hat Z}^2}{16\pi^4}s_{\hat \theta}^2c_{\hat \theta}^2\Big(1-\tfrac{m_{\hat Z}^2}{\mphi ^2}\Big)^3\; \hat c_{\hat Z\hat \gamma}^2, 
\end{align}
where the $c$ couplings are defined as,
\begin{align}
c_{g}&=\Big\vert b_{3}^{\rm eff}+\tfrac12 F_{1/2}(\tau_i) \Big\vert \; g_{\phi }\,,		&\hat c_{\hat g}&=\Big\vert \hat b_{3}^{\rm eff}+\tfrac12 F_{1/2}(\tau_{\hat i}) \Big\vert \; \hat g_{\phi }\,,	\\
c_{\gamma}&=\Big\vert b_{\textsc{em}}^{\rm eff}+\big(e_i^2N_c^i  F_{1/2}(\tau_i) \!-\!F_{1}(\tau_i)\big) \Big\vert\;g_{\phi }\,, 	
&\hat c_{\hat \gamma}&=\Big\vert \hat b_{\textsc{em}}^{\rm eff}+\big(e_{\hat i}^2 N_c^{\hat i}  F_{1/2}(\tau_{\hat i}) \!-\!F_{1}(\tau_{\hat i})\big) \Big\vert\;\hat g_{\phi }\,,	\\
c_{Z\gamma}&=\Big\vert b_{Z\gamma}^{\rm eff}+ \tfrac12 \big(\! A_{1/2}(\tau_i) \!+\!A_{1}(\tau_i) \big)\Big\vert\; g_{\phi }\,,	
&\hat c_{\hat Z\hat\gamma}&=\Big\vert \hat b_{\hat Z\hat \gamma}^{\rm eff}+\tfrac12 \big(\! A_{1/2}(\tau_{\hat i}) \!+\!A_{1}(\tau_{\hat i}) \big)\Big\vert\; \hat g_{\phi }\,.
\end{align}
The effective beta-function coefficients are defined in Eq.~\eqref{eq:bfun} and their explicit values are given below. 
The loop functions $F_{1/2}(\tau_i)$, $F_1(\tau_i)$, $A_{1/2}(\tau_i)$ and $A_{1}(\tau_i)$ are given in Appendix~\ref{sec:appendix}, and $\tau_i\equiv4m_i^2/m_\phi^2$ with the index $i$ denoting the particles in the corresponding loop. 
The dilaton partial decay width to a pair of Higgs bosons is,
\beq
\Gamma^{\phi }_{hh}=\frac{g_{\phi hh}^2}{32\pi}\frac1\mphi  \sqrt{1-\frac{4m_h^2}{\mphi ^2}}~, 
\eeq
where the trilinear coupling $g_{\phi hh}$ is given by,
\begin{align}
g_{\phi h h}&=\frac{m_{\phi }^2}{{\cal F}  } \Big(1+2 \frac{m_h^2}{m_{\phi }^2}\Big)\,.	\label{eq:gphh}
\end{align}

There are two types of contributions in the dilaton couplings to the massless gauge bosons. 
The first type of contributions arise from diagrams containing loops of SM or twin sector particles, these are included in the expressions above with the loop factors.
The second type of contributions are purely due to the running of the gauge couplings which are parameterized by the beta-function coefficients $b_i$, defined as $\beta_i(g_i)\!=\!b_i g_i^3/(16\pi^2)$.
These effective coefficients $b_i^{\rm eff}$ are parameterized as $b_i^{\rm eff}\!\equiv\! b_i^{\text{\tiny <}}-b_i^{\text{\tiny >}}$, where the $b_i^{\text{\tiny <}}$ and $b_i^{\text{\tiny >}}$ correspond to the IR and UV contributions to the beta-functions for the energy scales $0\!\leq\!\mu\!\leq\!\Lambda_\ir$ and $\Lambda_\ir\!\leq\!\mu\!\leq\!\Lambda_\uv$, respectively, and $\Lambda_{\ir(\uv)}$ is the IR (UV) cutoff. 
In the mirror twin Higgs model the gauge couplings are assumed to be equal at low energy,  and we speculate that this symmetry remains intact up to the UV with the gauge couplings running the same in both sectors. 
The reason that these coefficients are the same in the SM and twin sectors can be traced back to the $\mathbb{Z}_2$ symmetry relating the masses and couplings of the SM and twin states. 
Hence, in the MTH model the beta-function coefficients of the two sectors are the same, i.e. $\hat b_i\!=\!b_i$. 
However, the values of these $b_i^{\text{\tiny <}}$ and $b_i^{\text{\tiny >}}$ coefficients of the beta functions are highly model dependent as they depend on the number of states contributing to the gauge coupling running due to the IR and UV dynamics. 
In our convention, we consider the IR contributions to the $b_i^{\text{\tiny <}}$ terms as arising from all of the low energy SM and twin states, regardless of their degree of compositeness. Hence the $b_i^{\text{\tiny <}}$ in this model is equal to their corresponding SM values, i.e.
\beq
b_3^{\text{\tiny <}}=\hat{b}_{3}^{\text{\tiny <}}=-7\,, 		\lsp  b_2^{\text{\tiny <}}=\hat b_2^{\text{\tiny <}}=-\frac{19}{6}\,,		\lsp  b_1^{\text{\tiny <}}=\hat b_1^{\text{\tiny <}}=\frac{41}{6}\,.		\label{eq:bism}
\eeq

Furthermore, in this work we assume only the right-handed top quark and all the Goldstone bosons (including SM Higgs doublet) of the two sectors are composite, i.e. localized on the IR brane.
Whereas, all the remaining SM and their twin sector states are assumed to be partially composite, i.e. localized on the UV brane or in the bulk.
The fields localized on the UV brane or in the bulk contribute to the $b_i^{\text{\tiny >}}$ terms, and as well as the contributions from bulk gauge kinetic terms proportional to volume factor $1\!/\!(kL)$. 
What is unknown in these composite models are the UV sources $b_\textsc{cft}^{\text{\tiny >}}$ associated with the dynamics of the CFT breaking contributing to $b_i^{\text{\tiny >}}$. 
Therefore, in the following we consider two benchmark cases for the contribution due to the UV dynamics:
\bit 
\item[\underline{Case-A}:] 
We assume that the CFT dynamics cancel the contributions from the partially composite states and the bulk contributions proportional to $1/kL$, such that the total UV beta-function coefficients are zero, i.e. $b_i^{\text{\tiny >}}=\hat  b_i^{\text{\tiny >}}=0$. Hence, in this case, the total contribution to the effective $b$-coefficients $b_i^{\rm eff}\!\equiv\! b_i^{\text{\tiny <}}-b_i^{\text{\tiny >}}$ are
\beq
b_3^{\rm eff}=-7\,, 		\lsp  b_2^{\rm eff}=-\frac{19}{6}\,, 		\lsp  b_1^{\rm eff}=\frac{41}{6}\,.		\label{eq:caseA}
\eeq
\item[\underline{Case-B}:] In this case we assume that the UV contributions from the CFT dynamics are zero (or negligible), i.e. $b_\textsc{cft}^{\text{\tiny >}}=\hat  b_\textsc{cft}^{\text{\tiny >}}=0$. 
Therefore, the total UV contributions then arise solely from the running of the gauge couplings in both sectors due to the partially composite fields (bulk or UV-brane localized) and the bulk contribution due to volume factor, i.e. $b_{3}^{\text{\tiny >}}=\!-22/3+2\pi/(\alpha_\textsc{qcd}kL)$, $b_2^{\text{\tiny >}}=-10/3+2\pi/(\alpha_\textsc{em}kL)$, and $b_1^{\text{\tiny >}}=52/9+2\pi/(\alpha_\textsc{em}kL)$. 
Hence the total effective $b$-coefficients $b_i^{\rm eff}$ using Eq.~\eqref{eq:bism} are
\beq
b_3^{\rm eff}=\frac13-\frac{2\pi}{\alpha_\textsc{qcd}kL}\,, 		\lsp  b_2^{\rm eff}=\frac{1}{6}-\frac{2\pi}{\alpha_\textsc{em}kL}\,, 		\lsp  b_1^{\rm eff}=\frac{19}{18}-\frac{2\pi}{\alpha_\textsc{em}kL}\,.		\label{eq:caseB}
\eeq
Note the bulk contribution is proportional to the inverse of the 5D volume $kL$ which arises from the integration over the 5D bulk profiles of the gauge fields. 
\eit 

\begin{figure}[t!]
\centering
\includegraphics[width=0.5\textwidth]{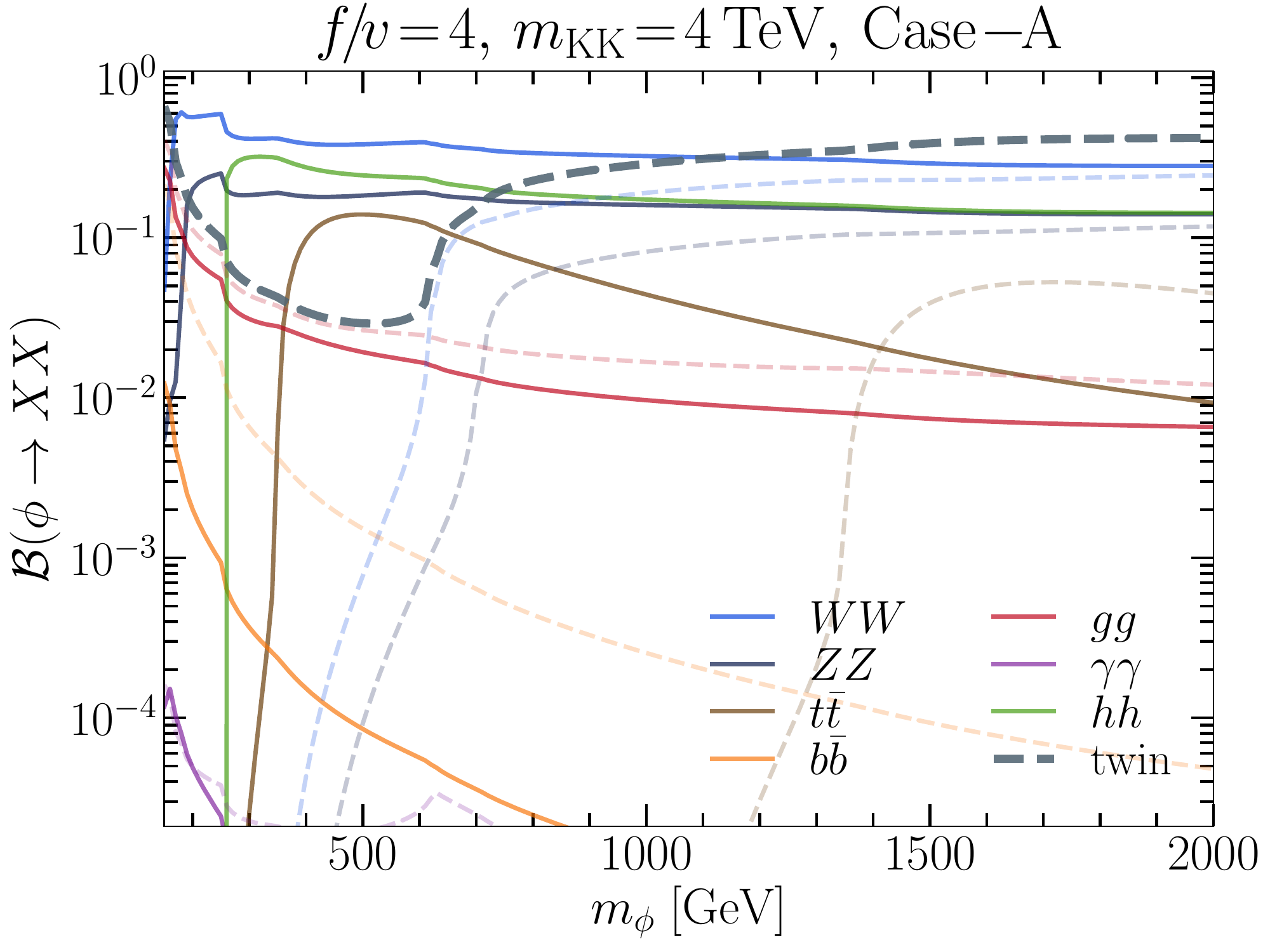}\includegraphics[width=0.5\textwidth]{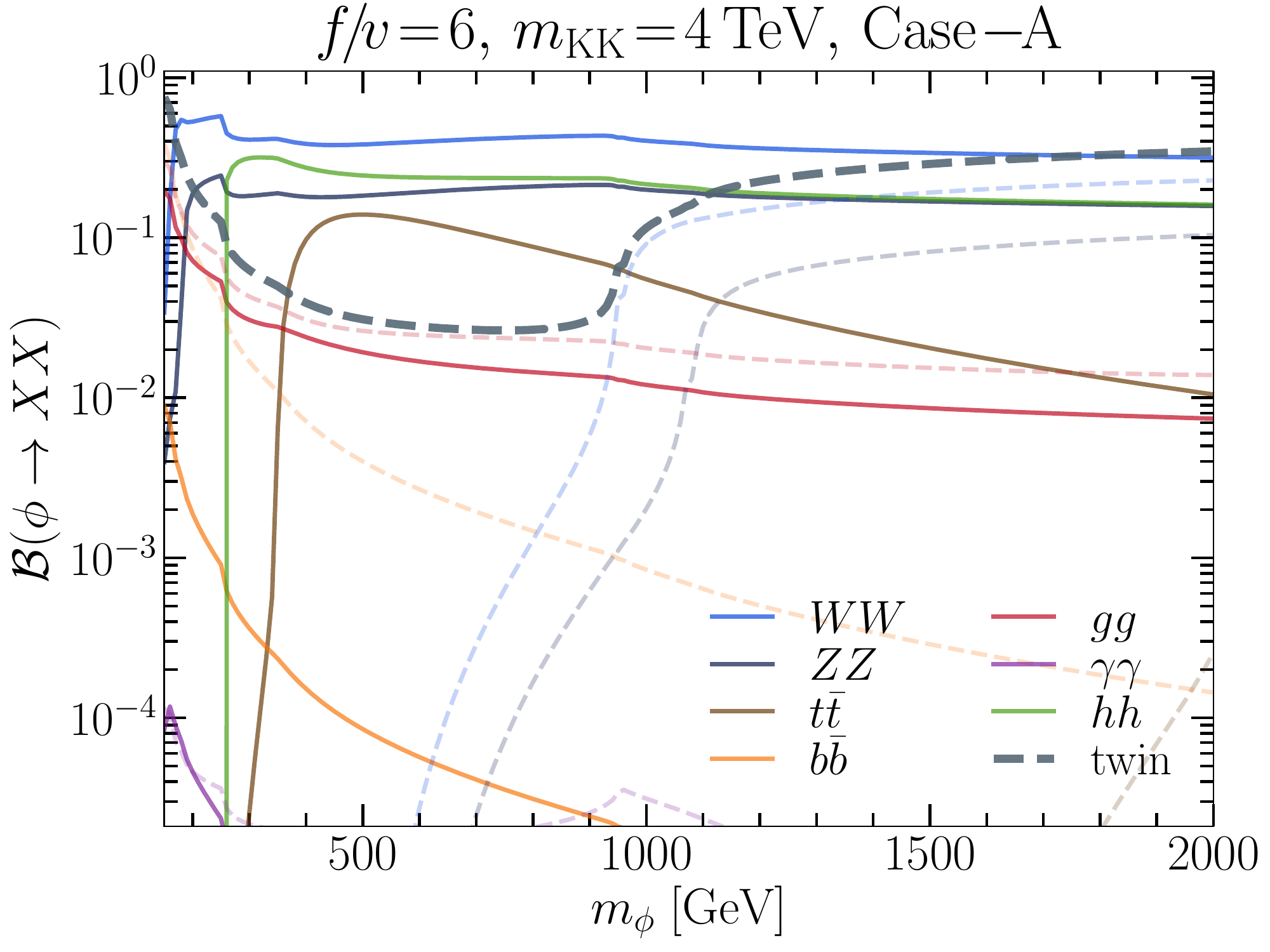}
\caption{Branching ratios of dilaton to the SM and twin sectors as a function of its mass for the choice of parameters $f\!/\!v \!=\!4$(left-panel), 6(right-panel) and $\mkk=4\tev$. For the anomalous contributions to massless gauge bosons we take values of $b$-coefficients for case-A. The solid curves represent the SM branching fractions, whereas, the corresponding twin curves are dashed (faded) and are summed into the gray dashed curve.}
\label{fig:brs}
\end{figure}

The branching ratios for the dilaton state to the SM and the twin sector as a function of dilaton mass are shown in Fig.~\ref{fig:brs}, for the parameter choices $f \!/\! v= 4$ (left-panel)  and $f \!/\! v= 6$ (right-panel), and $\mkk=4\tev$. Note that the dilaton branching fractions are independent of scale ${\cal F}$. The massless gauge boson branching ratios are those for the values of b-coefficients in case-A.
The analogous plots for case-B are similar.
Without invoking too much fine-tuning on the parameters of the effective potential for the dilaton, we assume in this work a dilaton mass in the range $[200\!-\!2000]\gev$. Such an assumption can be made based on estimates of the dilaton/radion mass from 5D RS-like models with GW stabilization mechanism, see e.g.~\cite{Csaki:2000zn,Chacko:2013dra}, where a radion mass $m_\phi\approx \op(0.1)\mkk$ is obtained with natural choices of 5D parameters. Furthermore, this dilaton mass range is also phenomenologically motivated by the fact that LHC experiments are probing such masses with reasonably high precision.  

\begin{figure}[t!]
\centering
\includegraphics[width=0.5\textwidth]{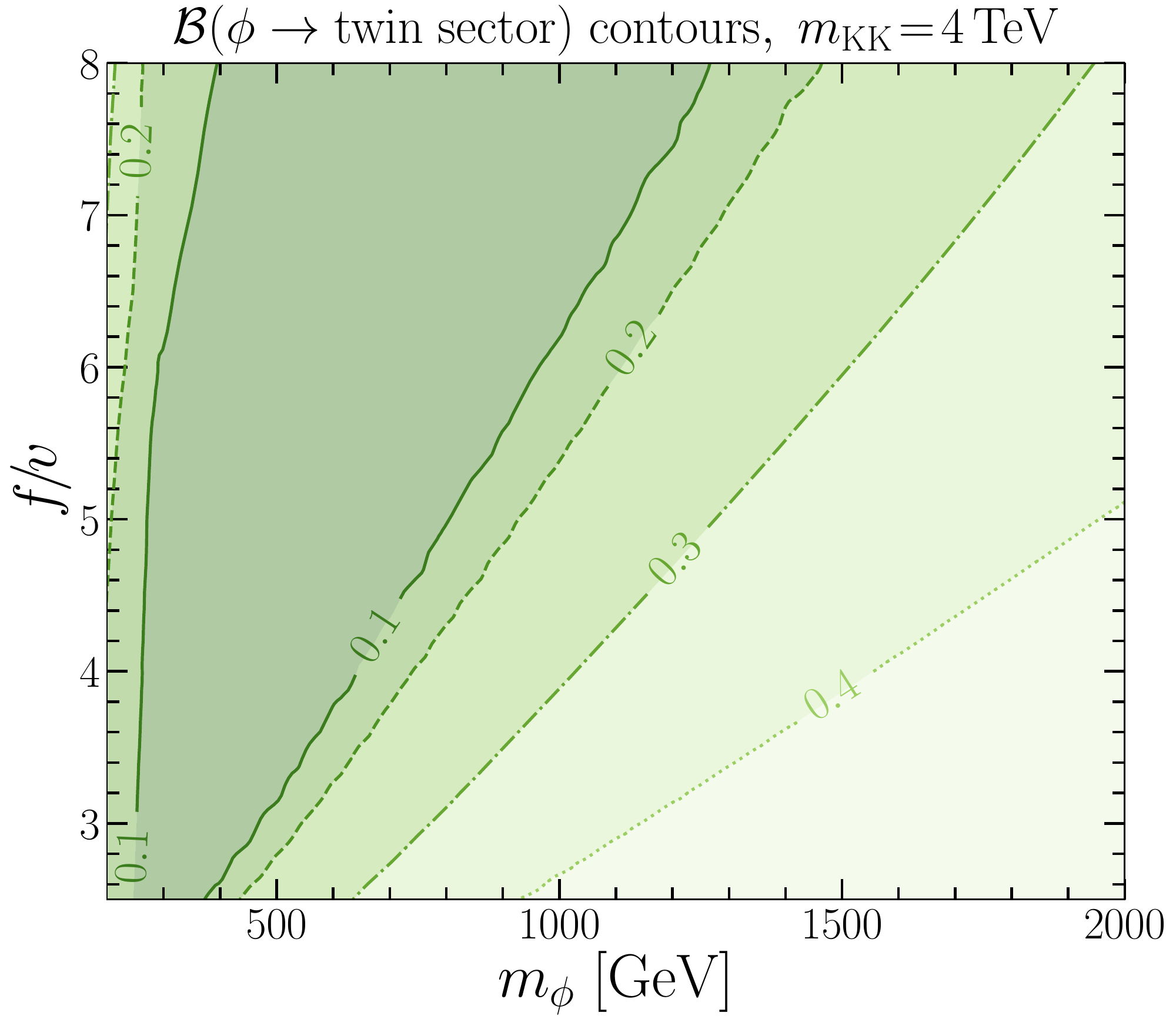}
\caption{The twin sector (invisible) branching fraction of the dilaton as a function of $m_\phi$ and $f/v$.}
\label{fig:invw}
\end{figure}

The important lessons from the branching ratios in Fig.~\ref{fig:brs} are as follows:
\bit
\item In the mass range below~$[200-650]\gev $ the dominant BR channels for the dilaton/radion are the SM gauge and Higgs bosons $WW/ZZ/hh$.
\item When the twin gauge bosons $\hat W\hat W/\hat Z\hat Z$ channels are allowed their BRs are approximately equal to those of the SM.
\item To a good approximation the following relation between the BRs holds for large dilaton masses ($m_\phi\!\geq\!2m_{\hat V}$)\,
\beq
{\cal B}(\phi\to hh)\simeq{\cal B}(\phi\to ZZ)\simeq\tfrac12{\cal B}(\phi\to WW)\simeq{\cal B}(\phi\to \hat Z\hat Z)\simeq\tfrac12{\cal B}(\phi\to \hat W\hat W)\simeq\frac17,
\eeq
which is a manifestation of Goldstone equivalence theorem. One can understand this result as originating from the fact that the SM Higgs boson and the longitudinal components of weak gauge bosons from the SM and twin sectors are all components of the same scalar field, i.e. they are all Goldstone bosons of the same spontaneous symmetry breaking. Furthermore, the apparent mass dependence of gauge boson couplings (see Feynman rules in Fig.~\ref{fig:feynrules}) are removed when we take into account the fact that the longitudinally polarized vector wave functions are proportional to $p^\mu/m_V$. 
\item The dilaton branching fractions to the massless gauge bosons in the both sectors are enhanced due to the trace anomalous and bulk contributions. 
\item The dilaton branching fractions to the fermions are proportional to the fermion masses, therefore the branching fractions to the twin sector fermions are larger by a factor $f/v$, which can be seen from the above figure in the bottom quark channels in the two sectors.
\eit
We note the dilaton branching fractions are independent of $\mathcal{F}$ and only weakly dependent on $\mkk$, which we have fixed to $4$ TeV. However, there could be a non-trivial dependence of its branching fractions on $f/v$, as the twin sector masses are directly proportional to this ratio.
To illustrate these effects we display contours of the invisible (twin sector) branching ratio of the dilaton as a function of $f/v$ and $m_\phi$ in Fig.~\ref{fig:invw} with $\mkk=4\tev$ and $b$-coefficients of case-A~\eqref{eq:caseA}. Note that the dilaton branching ratios to massless gauge bosons are subdominant, therefore a change in $b$-coefficients would not be significant for Fig.~\ref{fig:invw}.
We see that for larger values of $m_\phi$ more of the twin sector states become on-shell and the invisible branching ratio increases towards $\sim\!0.5$.
This occurs quicker for smaller values of $f$ since the twin sector states are then lighter.
%

\subsubsection*{Production cross sections at the LHC}
\label{crossection}
To perform the phenomenological analysis for the LHC constraints we scan over the dilaton portal parameters $m_\phi$ and $\cF$. Apart from $m_\phi$ and $\cF$, we have $f$ and $\mkk$ as the only other free parameters of the model. Note that fixing $f, \cF$ and $\mkk$ fixes $k/\mpl$ and $kL$ as these are related by Eqs.~\eqref{eq:mkk}-\eqref{eq:cF}.
Hence by fixing the parameters ($m_\phi, \cF, f,\mkk$) we can fully determine the phenomenology of the model up to the choice of $b$-coefficients for the massless gauge bosons. 
Therefore in this sense the model is very predictive. 

\begin{figure}[t!]
\centering
\includegraphics[width=0.5\textwidth]{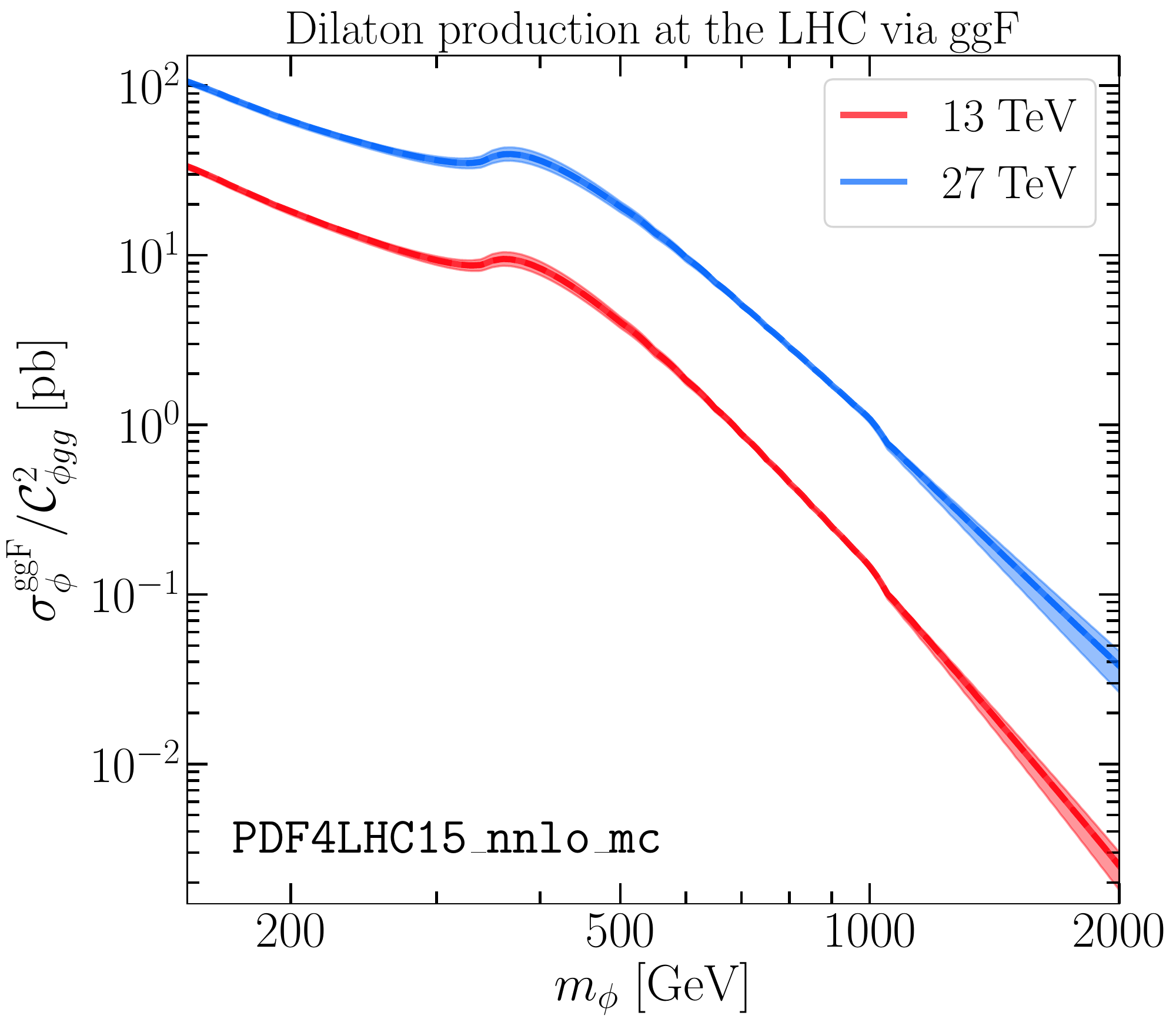}
\caption{This plot shows the dilaton production cross section normalized by ${\cal C}^2_{\phi gg}$ at the LHC via ggF for center of mass energy $13\tev$ and $27\tev$.}
\label{fig:xsec}
\end{figure}
We calculate the production cross section of the dilaton at the LHC, adopting a strategy similar to that outlined in~\cite{Ahmed:2015uqt}. The total cross-section, production$\times$branching fraction, of the dilaton is parameterized for different initial ($i$) and final ($j$) states as: 
\beq
\sigma^{i}_{\phi\to j}\!\equiv\!\sigma(i\to\phi)\!\cdot\!{\cal B}(\phi\to j)=\sigma^{\text{SM}}(i\to h)\big\vert_{m_h\!=\!m_{\phi}}\!\cdot{\cal C}_{\phi i}^2  \!\cdot\!{\cal B}(\phi\to j),
\eeq
where $\sigma^{\text{SM}}(i\!\to\! h)$ is the SM-like Higgs boson production cross-section calculated at the dilaton mass for $i\!=\!(gg,VV)$.  We have calculated the SM-like Higgs cross-sections via gluon-gluon fusion (ggF) at the dilaton mass, taking into account ${\rm N}^3{\rm LO}$ QCD and ${\rm NNLO}$ EW corrections by adoption of the fortran code \texttt{SusHi}~\cite{Harlander:2012pb,Harlander:2016hcx} with \texttt{PDF4LHC15\_nnlo\_mc} parton distribution functions and renormalization/factorization scales chosen to be $\mu\sim m_\phi/2$. 
The factor ${\cal B}(\phi\to j)$ is the dilaton branching fraction to the $j$ final state, whereas, effective coupling ${\cal C}_{\phi i}$ is defined as, 
\beq
{\cal C}^2_{\phi i} \equiv \frac{\sigma(i \to\phi)}{\sigma^{\text{SM}}(i \to h)|_{m_h\!=\!m_{\phi}}}=\frac{\Gamma(\phi\to i )}{\Gamma^{\text{SM}}(h\to i )|_{m_h\!=\!m_{\phi}}}~,
\eeq
where $\Gamma^{\text{SM}}(h\to i )|_{m_h\!=\!m_{\phi}}$ is the SM Higgs partial decay width evaluated at the dilaton mass and $\Gamma(\phi\to i )$ is the dilaton decay width in our model. 
In Fig.~\ref{fig:xsec} we show the dilaton cross sections $\sigma_\phi^{\rm ggF}$ normalized over ${\cal C}^2_{\phi gg}$ at the LHC via the gluon-gluon fusion (ggF) with center of mass energy $13 \tev$ and $27\tev$ (HE-LHC).
Note that the main production channel for the dilaton is ${\rm ggF}$, with vector-boson fusion (${\rm VBF}$) being subdominant in the parameter space considered. Since ggF is the most important production channel for the dilaton at the LHC, it is instructive to give the explicit expression for ${\cal C}^2_{\phi gg}$, i.e.
\beq
{\cal C}^2_{\phi gg}=\left\vert 1+ \frac{2b_{3}^{\rm eff}}{F_{1/2}(\tau_t)} \right\vert^2 \; \frac{v^2}{\cF^2}\,.		\label{eq:cphigg}
\eeq

\subsubsection*{Experimental bounds and projections}
\label{sec:bounds}
\begin{figure}[t!]
\centering
\includegraphics[width=0.5\textwidth]{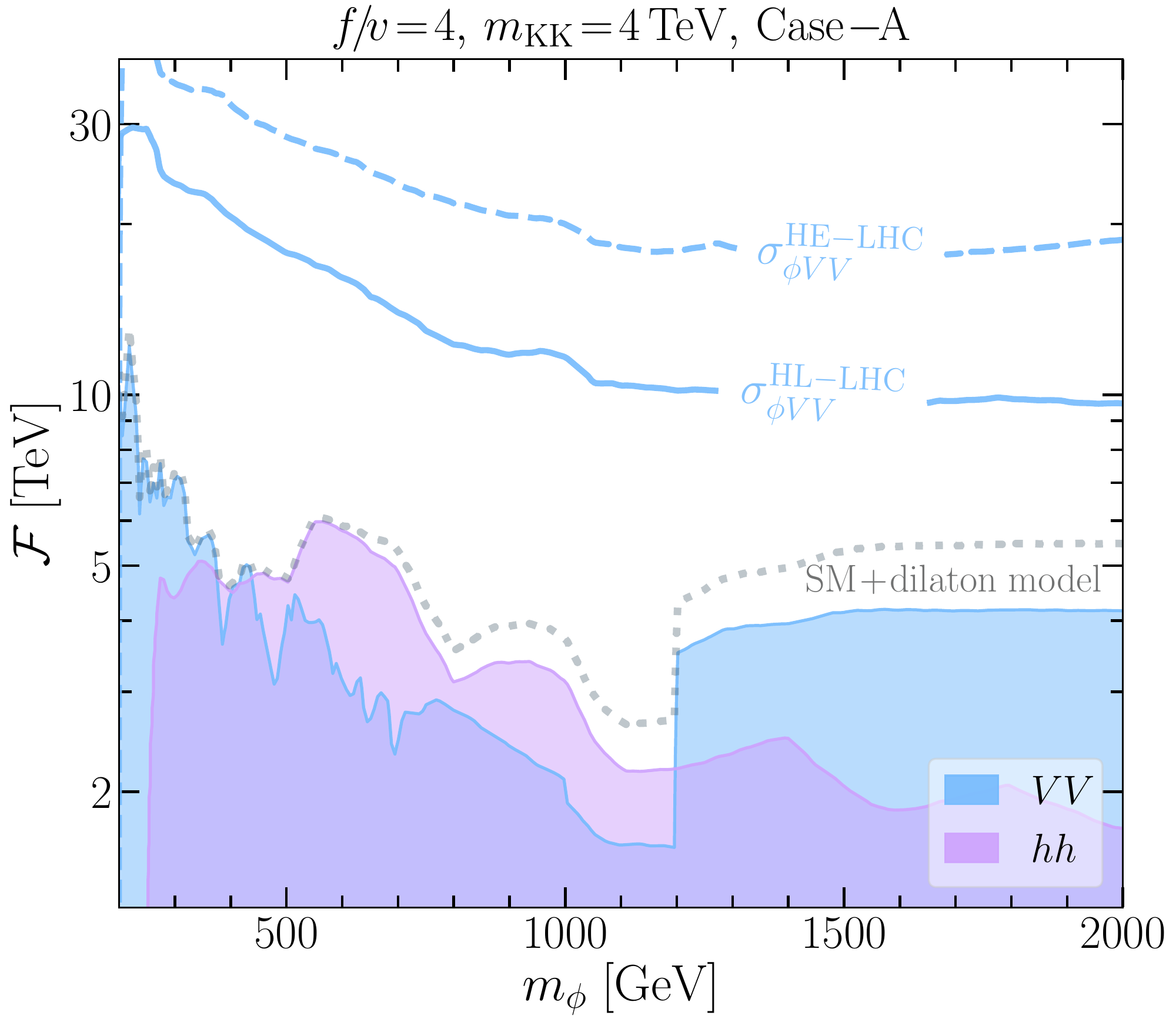}\includegraphics[width=0.5\textwidth]{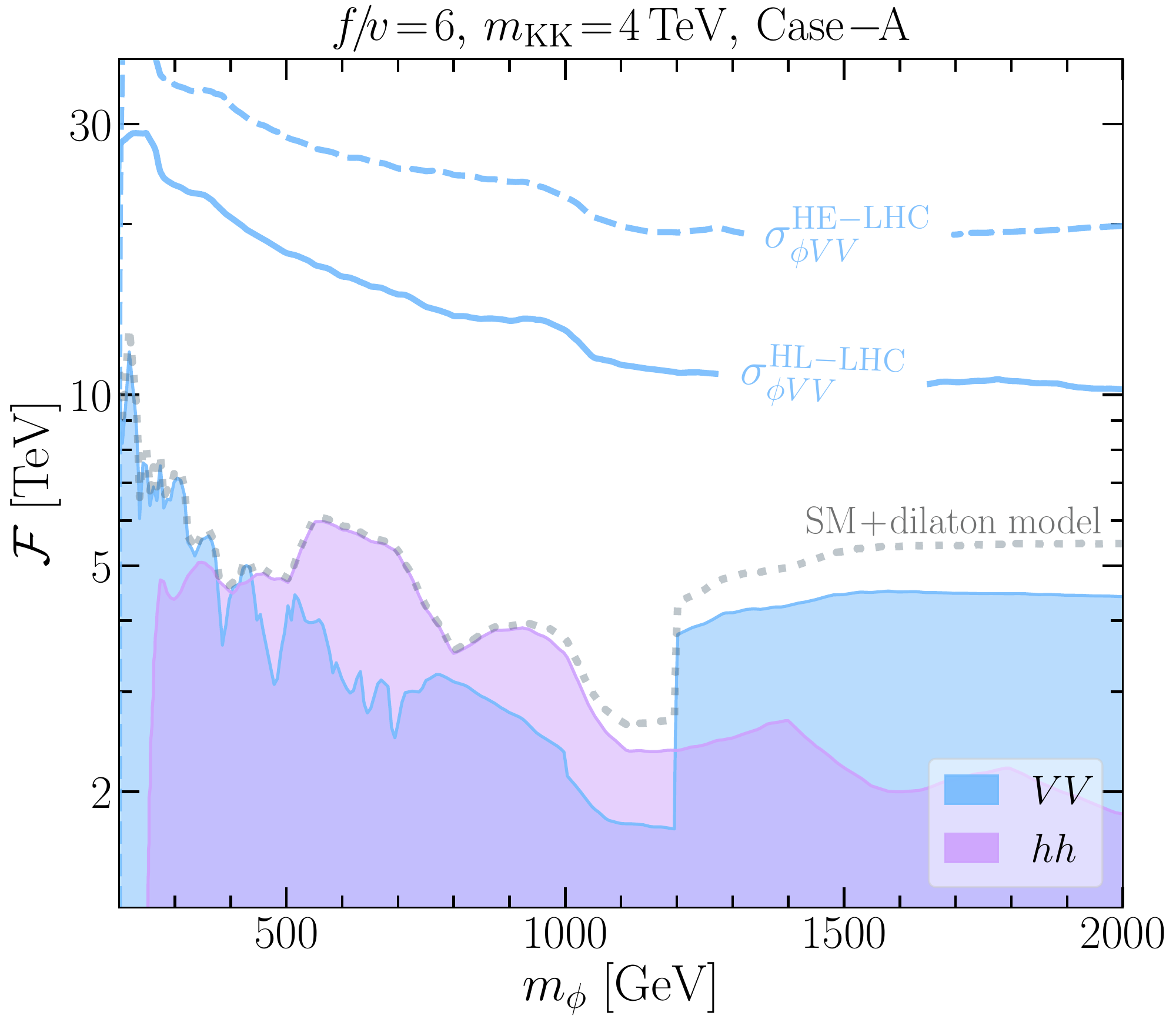}\\
\includegraphics[width=0.5\textwidth]{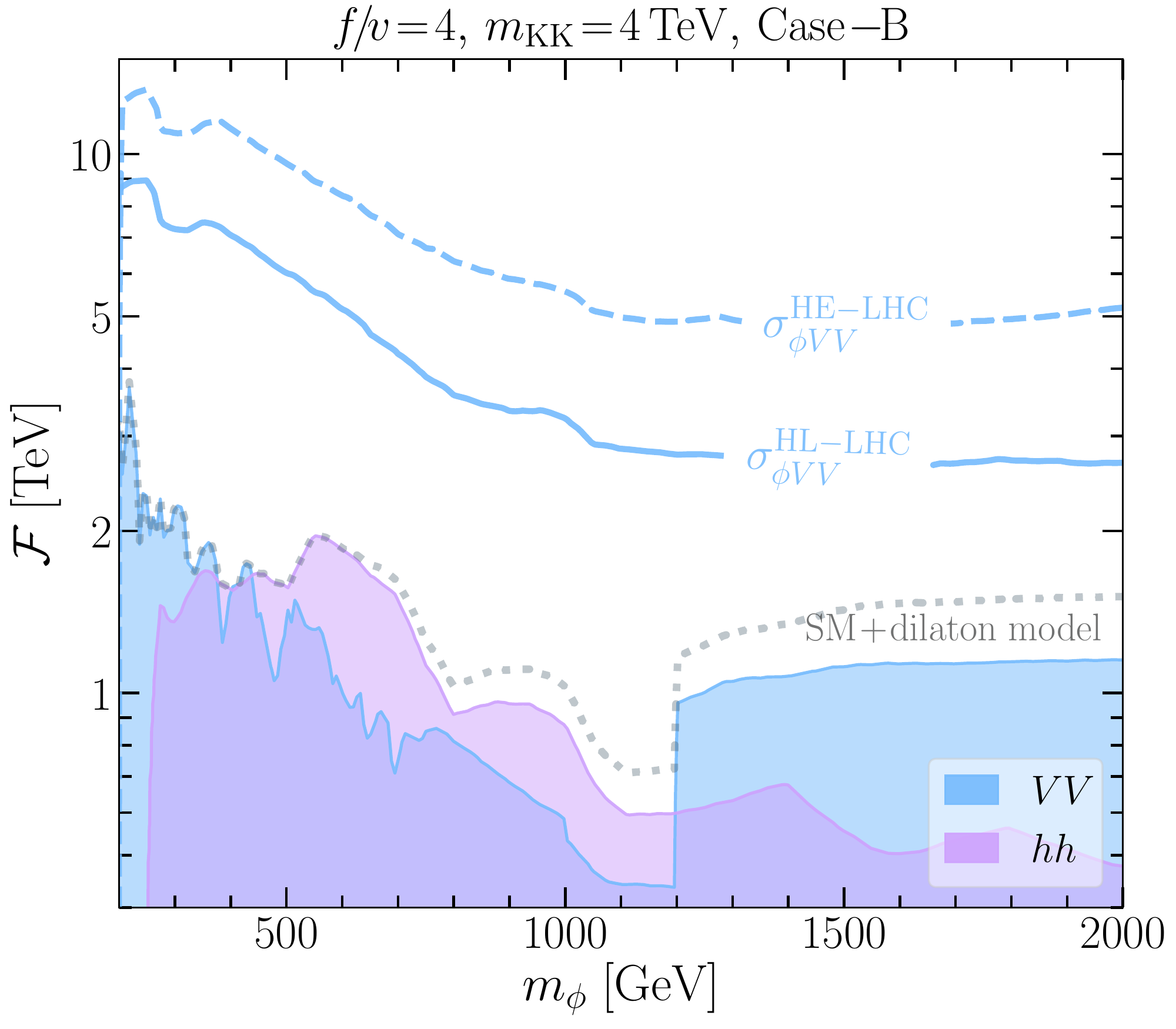}\includegraphics[width=0.5\textwidth]{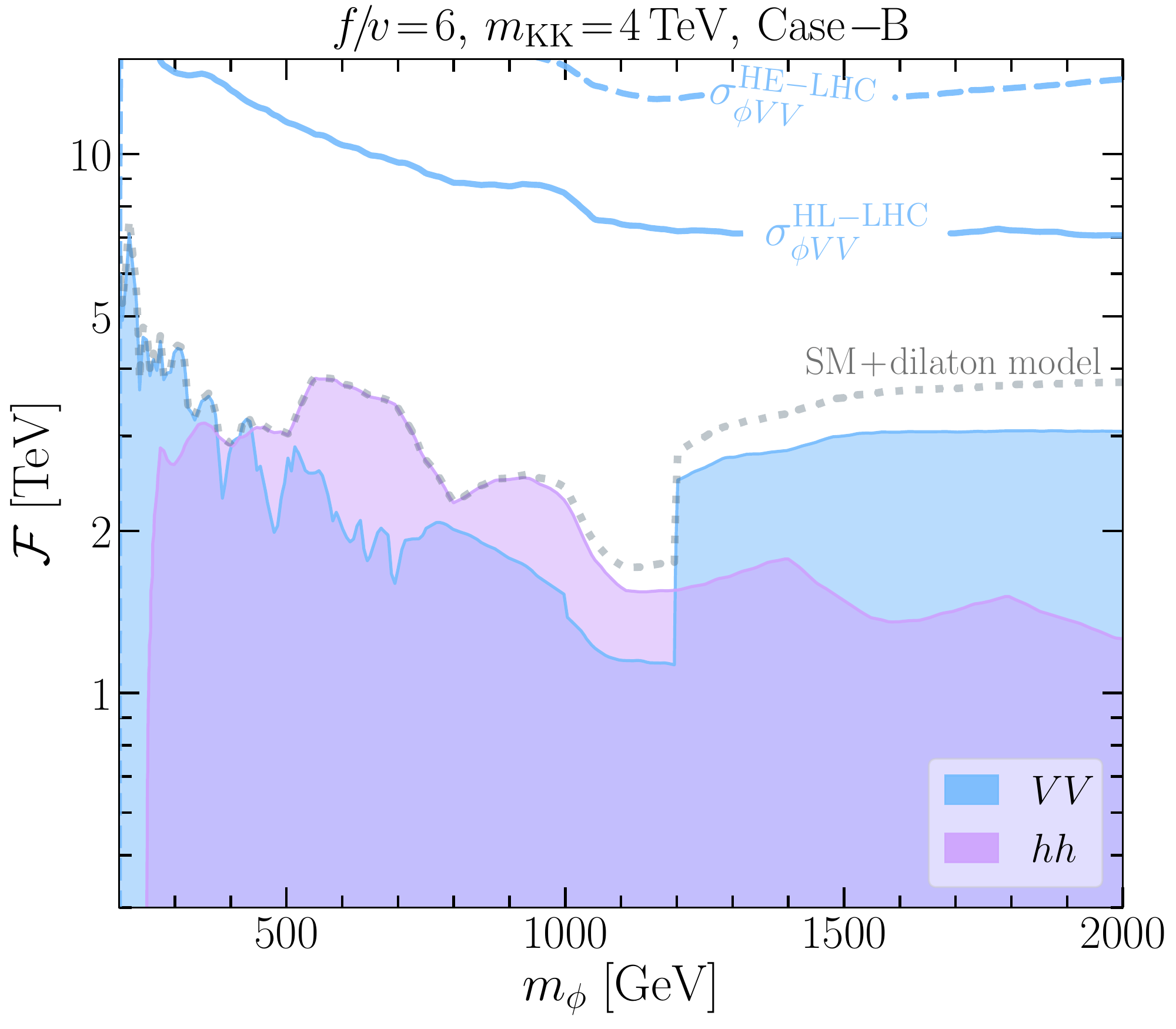}
\caption{The shaded regions show current LHC exclusion bounds from the dilaton cross sections to $VV$ (blue) and $hh$ (purple) on the VEV of dilaton ${\cal F} $ as a function of dilaton/radion mass for fixed values of $f /v=4$ and $f/v=6$, with $\mkk=4$ TeV.
The solid and dashed curves correspond to the HL-LHC ($\sqrt{s}\!=\!14\tev$) and HE-LHC ($\sqrt{s}\!=\!27\tev$) reach at 95\% C.L. with integrated luminosity $3000~{\rm fb}^{-1}$, respectively. }
\label{fig:bounds}
\end{figure}

The exclusion bounds in the $m_\phi\!-\!\cF$ plane for fixed values of $f\!/\!v=4$  (left-panel) and $f\!/\!v=6$ (right-panel), with $\mkk=4$ TeV are shown in Fig.~\ref{fig:bounds}. The upper and lower plots are for effective $b$-coefficients values of case-A and case-B, respectively. The shaded region is excluded by the LHC run-2 data from the ATLAS and CMS experiments. The experimental constraints employed in the our analysis are from Refs.~\cite{CMS:2019xud,CMS:2019kjn,Sirunyan:2019jbg,Aaboud:2018bun,ATLAS:2016oum,CMS:2016ilx,Sirunyan:2019vgt} and 
\cite{Aaboud:2018ewm,Aaboud:2018sfw,Sirunyan:2018two,Aad:2019uzh,Sirunyan:2019quj} for $VV\!=\!WW,ZZ$ and $hh$, respectively. The blue (purple) shaded region represents the exclusion of dilaton to $VV$ ($hh$) final state, where $V\!=\!W,Z$. For the effective $b$-coefficients in case-A (upper-panel), the present exclusion bound on the scale of conformal breaking $\cF$ of $\sim\!\op(5\!-\!8)\tev$ for dilaton masses $m_\phi\!=\![200-2000]\gev$ is mainly due to $VV$ and $hh$ final states, which is to be expected as these states have the largest branching fractions ${\cal B}(\phi\to hh)\simeq{\cal B}(\phi\to VV)\simeq1/7$, see Fig.~\ref{fig:brs}. 
The constraints in all other channels are much weaker as compared to that of the di-boson final states.

In the lower-panel of Fig.~\ref{fig:bounds}, we consider case-B for the values of the effective beta function coefficients where the UV contributions are zero and the only contributions are due to the composite states in the IR. In particular, the QCD beta function coefficient is $b_{3}^{\rm eff}\!=\!1/3-2\pi/(\alpha_\textsc{qcd}kL)$ which is much smaller for $kL\sim35$ than the one considered in case-A, where its value was $-7$. Going from case-A to B, the dilaton effective coupling ${\cal C}_{\phi gg}$ decreases by a factor of $\op(3)$ for fixed $\cF$. This strong reduction in the cross-sections leads to a weakening of the constraints on $\cF$ by the same amount as shown in Fig.~\ref{fig:bounds}. For  case-B (lower-panel), we see the constraint on $\cF$ is $\sim\op(1\!-\!3)\tev$ with $f/v\!=\!4\!-\!6$ in most of the dilaton mass range considered. 
Note by increasing the value of $f/v$ the constraints get stronger as the dilaton production cross section via ggF gets an enhancement. This non-trivial effect can be understood by noting that for a fixed $\mkk$ (as in our case), increasing $f$ reduces the value of $kL$ (as seen in Eq.~\eqref{eq:mkk}), which then enhances the dilaton production cross section through its dependence on the effective coupling ${\cal C}_{\phi gg}$~\eqref{eq:cphigg}. As we go from $f/v\!=\!4$ to $6$, the direct search constraints on $\cF$ get $\op(20\%)$ stronger. 
Since the bulk volume contribution to the $b^{\rm eff}_3$ is not present in the case-A~\eqref{eq:caseA} therefore, we do not observe strong enhancement in the dilaton production and as a result there is no significant change in the constraints on $\cF$ from changing $f\!/\!v:4\to6$, see upper-panels in Fig.~\ref{fig:bounds}.

In addition to the existing constraints, the solid and dashed blue curves in Fig.~\ref{fig:bounds} show the expected 95\% C.L. reach due to the dilaton decaying to $VV$ final states at the HL-LHC with $\sqrt{s}\!=\!14\tev$ and at the HE-LHC upgrade with $\sqrt{s}\!=\!27\tev$ with 3000 ${\rm fb}^{-1}$, respectively. The HL-LHC (HE-LHC) projected reach shows that the scale $\cF$ will be probed at $\sim\!10(20)\tev$ for dilaton masses up to $2\tev$ in case-A (upper-panel), whereas, in  case-B (lower-panel) the reach is weakened by a factor of $\op(3)$. 
Furthermore, in these plots the dotted gray curves show the constraints on the scale $\cF$ in the absence of twin sector, i.e. only the SM and the dilaton. We note that the presence of the twin sector weakens the constraints on the scale $\cF$ for relatively large dilaton masses, which is expected as for a large dilaton mass the twin gauge boson channels and twin top channels become kinematically accessible.

\subsection*{Global fit to the twin Higgs parameters}
\label{sec:globalfit}
In these models the mixing between the pseudo-Goldstone Higgs boson and the dilaton field is negligible, therefore, the Higgs boson behaves very similarly to how it behaves in the holographic/composite Higgs models. Except now there will be invisible decays of the Higgs to twin sector states provided that they are kinematically allowed. The pseudo-Goldstone Higgs field couplings to the SM and twin gauge bosons are given by
\begin{align}
g_{hVV}^{}=\sqrt{1-\tfrac{v^2}{f ^2}}\,g_{hVV^{}}^{\rm{SM}},\lsp 	g_{h\hat{V}\hat{V}}=-\sqrt{1-\tfrac{v^2}{f ^2}}\,g_{hVV^{}}^{\text{\tiny SM}},		\label{eq:ghvv}
\end{align}
where $g_{hVV^{}}^{\rm{SM}}\!=\!2m_V^2/v$.
The Yukawa couplings of the SM Higgs to the fermions in both sectors are slightly more involved, but can be calculated from the form factors in the effective Lagrangian.
They are found to be approximately
\beq
\begin{aligned}
g_{h\psi\psi}&=\partial_vm_f\Big|_{p_E\rightarrow0}\simeq\sqrt{1-\tfrac{v^2}{f ^2}}\,\frac{m_\psi}{v}=\sqrt{1-\tfrac{v^2}{f ^2}}\,g_{h\psi\psi^{}}^{\rm{SM}},	\\
g_{h\hat \psi\hat{\psi}}&=\partial_v m_{\hat{\psi}}\Big|_{p_E\rightarrow0}\simeq-\frac{v}{f }\frac{m_{\hat \psi}}{\hat v}=-\frac{v}{f }\,g_{h\psi\psi^{}}^{\rm{SM}},	
\end{aligned} \label{eq:ghff}
\eeq
where $g_{h\psi\psi^{}}^{\rm{SM}}\!=\!m_\psi/v$.
The loop induced couplings of the Higgs to gluons receives contributions primarily from the top quark and can be expressed simply as (see also Sec.~\ref{sec:holo_twin}),
\begin{align}
g_{hgg}\simeq&\sqrt{1-\tfrac{v^2}{f ^2}}\,g_{hgg}^{\text{\tiny SM}}
\end{align}
while the loop induced coupling to photons involves loops of both the top quark and $W$ bosons.
Through the couplings to the twin quarks the Higgs will also couple to the twin gluons and the twin photon.
For more details on the loop induced couplings of the pseudo-Goldstone Higgs we refer the reader to \cite{Falkowski:2007hz,Azatov:2011qy,Carena:2014ria}.

To analyse the compatibility of the SM-like Higgs boson in the composite/holographic twin Higgs models with the data from the LHC experiments, we calculate the cross-section times branching fractions as, 
\beq
\sigma^{i}_{h\to j}\!\equiv\!\sigma(i\to h)\!\cdot\!{\cal B}(h\to j)=\sigma^{\text{SM}}(i\to h)\cdot{\cal C}_{h i}^2  \cdot{\cal B}(h\to j),
\eeq
where for $i=(gg,VV)$ the effective SM Higgs couplings in the model are
\[
{\cal C}^2_{h gg} ={\cal C}^2_{h VV} = 1-\frac{v^2}{f ^2}\,.
\]
The SM Higgs signal-strength measurements and its couplings in many production and decay channels have been measured with impressive precision at the LHC Run-2 and they provide stringent constraints on BSM models. We perform a global $\chi^2$ fit to the Higgs signal strength measurements, defined as production$\times$branching fractions relative to those of the SM expectation, i.e.
\beq
\mu^{i}_{j}\equiv \frac{\sigma(i\to h)\!\cdot\!{\cal B}( h\to j)}{\sigma^{\text{\tiny SM}}(i\to h)\!\cdot\!{\cal B}^{\text{\tiny SM}}(h\to j)}={\cal C}_{h i}^2\,\frac{{\cal B}(h\to j)}{{\cal B}^{\text{\tiny SM}}(h\to j)}\,.	\label{eq:signalstrength}
\eeq 
This is performed with \texttt{Lilith-2}~\cite{Bernon:2015hsa,Kraml:2019sis} which incorporates all of the relevant LHC Run-2 data. For the global $\chi^2$ analysis \texttt{Lilith-2} employs variable Gaussian and Poisson likelihoods with full LHC Run-2 data from ATLAS and CMS for $36~{\rm fb}^{-1}$. For details on the implementation of \texttt{Lilith-2} and experimental data used, see Refs.~\cite{Bernon:2015hsa,Kraml:2019sis}.

\begin{figure}[t!]
\centering
\includegraphics[width=0.57\textwidth]{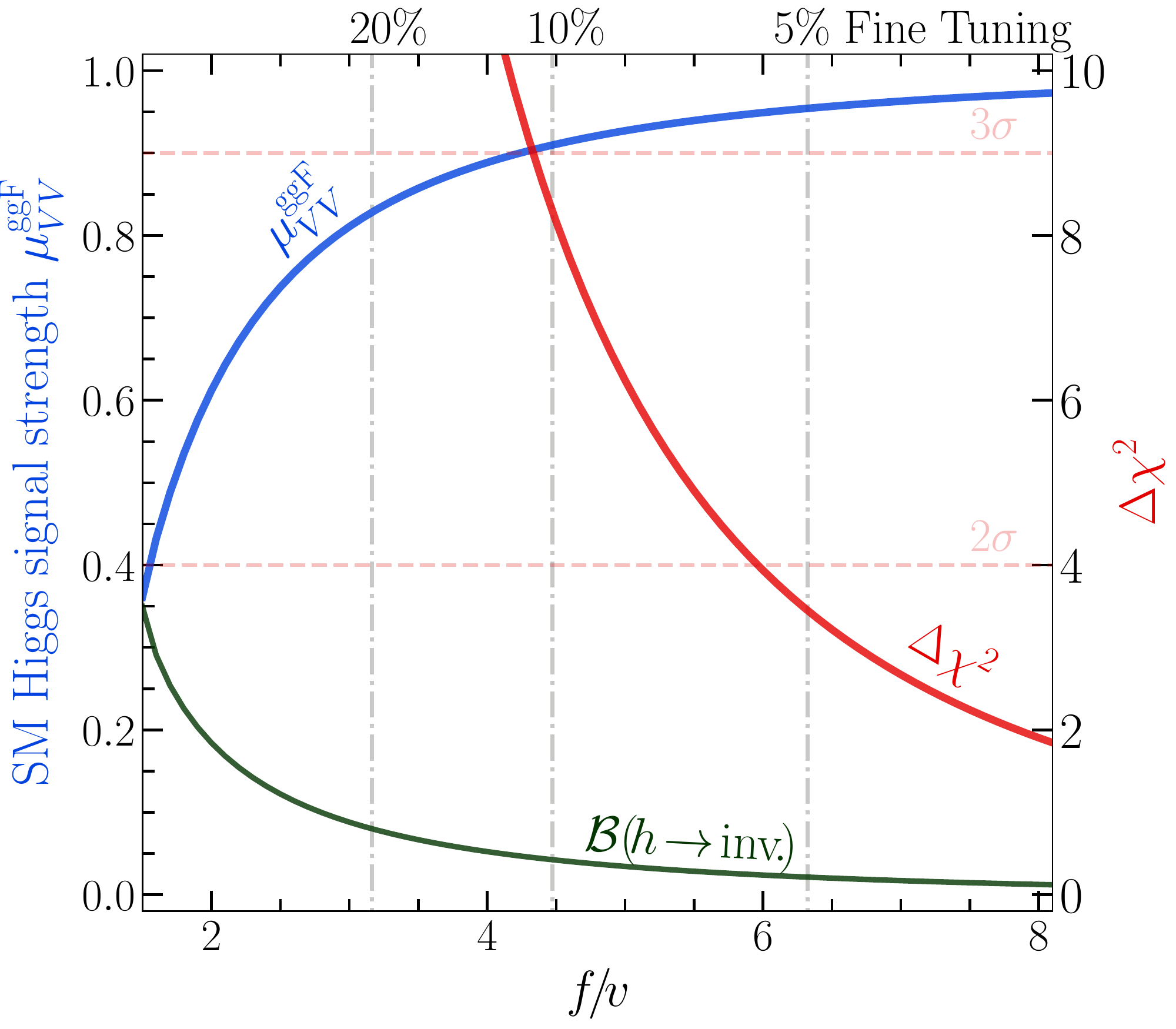}
\caption{In this plot we summarize our results for the global fit of the SM Higgs data from LHC Run-2 as a function of $f\!/\!v$. The blue curve shows the SM Higgs signal strength measurement $\mu_{VV}^{\rm ggF}$ with the markers on the left y-axis, and the red curve shows the $\Delta\chi^2$ fit values to the SM Higgs signal strength measurements with the markers on the right y-axis. The green curve shows the SM Higgs invisible branching fraction ${\cal B}(h\!\to\! {\rm inv}.)$, i.e. the SM Higgs decays to the twin sector (markers are w.r.t. left y-axis). We also show with dashed light-red $2\sigma$ and $3\sigma$ lines for the $\Delta\chi^2$ fit to the Higgs data. The dash-dotted vertical lines show the estimated fine tuning required in the twin Higgs model.}
\label{fig:higgsfit}
\end{figure}

In Fig.~\ref{fig:higgsfit} we summarize the constraints from the SM Higgs measurements on the model as a function of $f\!/\!v$.  Note that $f\!/\!v$ is the only parameter relevant for these measurements in this model. 
The blue curve shows the SM Higgs signal strength measurement $\mu_{VV}^{\rm ggF}$ where $V=W/Z$ with left y-axis markers. Whereas, the $\Delta\chi^2$ fit to the SM Higgs signal strength measurements is shown in red curve with markers on the right y-axis. 
The green curve shows SM Higgs invisible branching fraction ${\cal B}(h\!\to\! {\rm inv}.)$, i.e. the SM Higgs decays to the twin sector (markers are w.r.t. left y-axis). 
We also show with dashed light-red $2\sigma$ and $3\sigma$ lines for the $\Delta\chi^2$ fit to the Higgs data. Note that $\Delta\chi^2$ at $3\sigma$ already pushes $f\!/\!v\gtrsim 4$ which corresponds to about $10\%$ fine tuning. 
The gray dash-dotted vertical lines show fine tuning required in the twin Higgs model, i.e. $\Delta_{\textsc{ft}}\!\equiv\!2v^2\!/\!f^2$. 
It is projected that the Higgs signal strength measurements at the HL-LHC would push the scale further up to $f\!/\!v\gtrsim 6$ corresponding to a fine-tuning less than $5\%$.
%

\section{Conclusions}
\label{sec:conc}
Twin Higgs scenarios provide an enticing alternative to the more widely studied frameworks of minimal composite Higgs and supersymmetric models.
With their ability to elegantly explain a little hierarchy between the electroweak scale and a new physics scale, they are perhaps one of the best motivated new physics scenarios to probe at the LHC.
In this paper we consider a strongly interacting twin Higgs model, where the UV theory is approximate scale invariant. 
This model predicts a relatively light dilaton in the low energy EFT, which provides a portal between the SM and twin sectors. 
This opens up a new avenue in twin Higgs phenomenology, where we demonstrate that the presence of a twin sector can have dramatic effects on the phenomenology of a dilaton state with mass near TeV scale.

We started with a discussion of a UV completion of the twin Higgs mechanism through a strongly coupled scale invariant UV theory, and derived the low energy theory based on minimal global/scale symmetry conditions designed to produce the required matter structures.
The presence of approximate scale invariance in the UV leads to the introduction of a new degree of freedom in the IR, the dilaton field.
The fact that this field lives close to the compositeness scale introduces an even richer phenomenology for the new physics sector.
Based on symmetry and scaling arguments we derive the form of the interactions that this dilaton field should have within the SITH model.
The results thus far had been quite general, however a calculable framework in which to constrain the parameter space of these models is preferred, therefore we then introduced the 5D holographic twin Higgs model \cite{Geller:2014kta}.
The AdS/CFT correspondence tells us that these holographic models based on the RS-like geometry with a stabilization mechanism should mimic the effective theory of a strongly coupled scale invariant gauge theory which confines in the IR ($\sim$ TeV).
We detailed the construction of the twin Higgs model in this 5D framework, where the introduction of the dilaton field occurs naturally in the guise of the radion field.
Interactions between the dilaton/radion field and the states in the twin Higgs model are derived and approximate relations for the couplings and branching ratios are given.
We find that the parameter space of the model is covered by four parameters, $(m_\phi,\mathcal{F},f,\mkk)$, and from here we perform our analysis.

We begin the analysis with a study of the branching ratios and show the impact that the presence of the twin sector has.
As expected, for large dilaton mass the branching ratio to SM states reduces to roughly $50\%$ of what it would be were the twin sector not there, this can have a considerable effect on the radion phenomenology.
We proceed to compute the exclusion bounds (Fig.~\ref{fig:bounds}) on the scale $\mathcal{F}$ for differing values of $f/v$, demonstrating that the current bounds lay only the range $\mathcal{O}({\rm few}) \tev$, aided through the presence of the twin sector states.
In fact we see a significant effect on the exclusion bounds when going only from $f/v=4$ to $f/v=6$.
Furthermore, we illustrate the significance of IR and UV contributions to the running of gauge couplings in this model which define the dilaton couplings to massless gauge bosons. 
In the same plot we presented the projected reach of HL-LHC ($14\tev$) and HE-LHC ($27\tev$) with $3000$fb$^{-1}$ at $95$\% C.L. demonstrating that these LHC upgrades could take the bound on $\mathcal{F}$ to $\mathcal{O}$($10$) TeV even with the presence of the twin states.
Lastly we considered possible constraints on $f/v$ due to deviations in the Higgs boson's couplings  and its invisible decays, since in the presence of the twin sector the Higgs boson can now have an appreciable invisible decay width.  
Performing a global $\chi^2$ fit to the Higgs signal strength measurements we find that at $3\sigma$ we already require $f/v\gtrsim4$.

In this paper we have performed a detailed study of dilaton phenomenology in a composite twin Higgs model, using the holographic correspondence as a tool to provide a calculable framework in which to do so.
The results of the study not only provide the current bounds on the model parameters, but also demonstrate the power that increased precision on Higgs coupling measurements and increased sensitivity in resonance searches will have on the parameter constraints. It is straightforward to generalize our results to other neutral naturalness models with strongly coupled UV completions. 

\section*{Acknowledgements}
We thank Zackaria Chacko and Alberto Mariotti for useful discussions and comments.  
AA and SN are supported by FWO under the EOS-be.h project no. 30820817 and Vrije Universiteit Brussel through the Strategic Research Program ``High Energy Physics''.
BMD  acknowledges  the  financial  support  from  the  Slovenian Research Agency (research core funding No.  P1-0035 and J1-8137).

\appendix
\section{Feynman rules}
\label{sec:appendix}
\begin{figure}[t!]
\centering
\includegraphics[width=\textwidth]{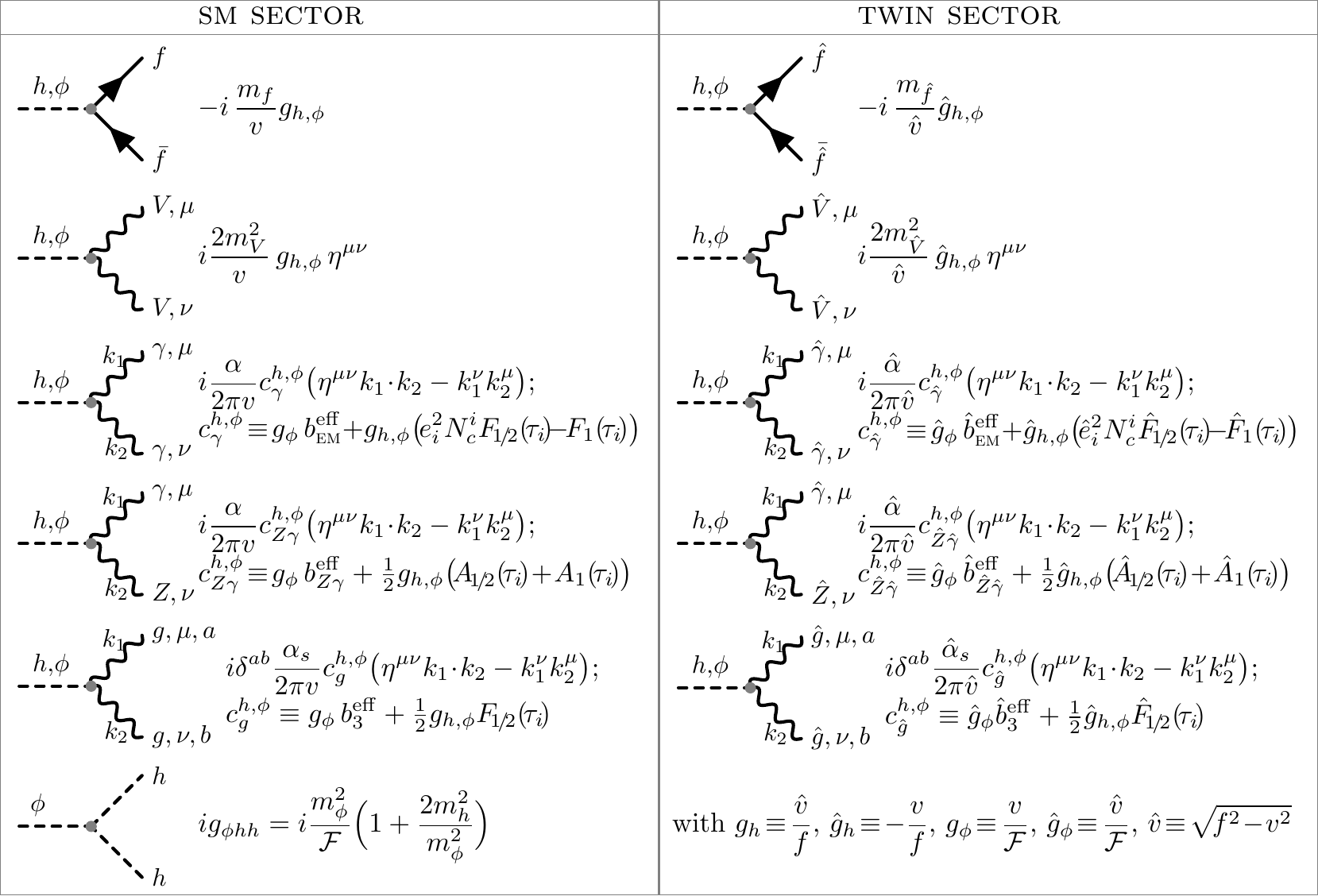}
\caption{Feynman rules for the SM and twin sector particles coupling with the SM Higgs $h(x)$ and the dilaton/radion $\phi(x)$ in the low energy effective theory of a dilaton portal composite twin Higgs model. In our notation the effective $b_i^{\rm eff}\!\equiv\!b_i^{\text{\tiny <}}-b_i^{\text{\tiny >}}$ are the coefficients of the gauge coupling beta functions defined as, $\beta_i(g_i)\!=\!b_i^{\rm eff} g_i^3/(16\pi^2)$, where $i\!=\!3,2,1$ corresponds to gauge couplings of $SU(3)_c, SU(2)_L$, and $U(1)_Y$, respectively, and are defined in Eq.~\eqref{eq:bfun}. The triangle loop functions are given in this appendix. The {\it hatted}\, quantities represent the twin sector.}
\label{fig:feynrules}
\end{figure}

In this Appendix we present the Feynman rules for the dilaton in a SITH model. We derived the Feynman rules in Fig.~\ref{fig:feynrules} under the assumption that corrections in the dilaton couplings to the massive fermions and gauge bosons of the SM and twin sectors due to the explicit breaking of CFT in the UV are negligible. This assumption is verified in both the CFT and holographic pictures in Ref.~\cite{Chacko:2012sy,Chacko:2014pqa}, see also~\cite{Chacko:2013dra}. It was shown in~\cite{Chacko:2012sy,Chacko:2014pqa} that these corrections are proportional to $m_\phi^2/\cF^2$ or $m_\phi^2/\mkk^2$, hence in the limit when the dilaton is relatively light, as in our case, one can neglect these corrections. 
Regarding the dilaton couplings to massless gauge bosons in the both sectors, in Sec.\ref{sec:pheno} we consider two cases for the UV contributions from the CFT breaking in the running of gauge couplings, i.e. the effective $b$-coefficients. In Fig.~\ref{fig:feynrules} we collect all the relevant three point vertices for the dilaton portal phenomenology in the composite twin Higgs models. 

Following the notations of the Higgs Hunter's Guide~\cite{Gunion:1989we}, the form factors $F_{1/2}$, $F_1$, $A_{1/2}$ and $A_1$ employed in the Feynman rules (Fig.~\ref{fig:feynrules}) are
\begin{align}
F_{1/2}(\tau_i)&=-2\tau_i\big[1+(1-\tau_i)f(\tau_i)\big], 
\lsp F_{1}(\tau_i)=2+3\tau_i+3\tau(2-\tau_i)f(\tau_i), \\
A_{1/2}(\tau_i,\kappa_i)&=\frac{-e_iN_c^i}{\sw\cw}\big(1-4e_i\swsq\big) \big[{\cal I}_1(\tau_i,\kappa_i)-{\cal I}_2(\tau_i,\kappa_i)\big]		,\\
A_1(\tau_i,\kappa_i)&=-\frac{\cw}{\sw}\Big[4\Big(3-\frac{\swsq}{\cwsq}\Big){\cal I}_2(\tau_i,\kappa_i)	\notag\\
&\hspace{2cm} + \Big((1+2/\tau_i)\frac{\swsq}{\cwsq}- (5+2/\tau_i)\Big){\cal I}_1(\tau_i,\kappa_i))\Big],		
\end{align}
where $\tau_i\equiv4m_i^2/\mphi ^2$, $\kappa_i\equiv 4m_i^2/m_Z^2$, and 
\begin{align}
{\cal I}_1(\tau_i,\kappa_i) &= \frac{\tau_i\kappa_i}{2(\tau_i-\kappa_i)}+ \frac{\tau_i^2\kappa_i^2}{2(\tau_i-\kappa_i)^2}\Big(f(\tau_i)-f(\kappa_i)\Big)+ \frac{\tau_i^2\kappa_i}{(\tau_i-\kappa_i)^2}\Big(g(\tau_i)-g(\kappa_i)\Big),	\\
{\cal I}_2(\tau_i,\kappa_i) &= -\frac{\tau_i\kappa_i}{2(\tau_i-\kappa_i)}\Big(f(\tau_i)-f(\kappa_i)\Big),	\\
f(\tau_i)&= \begin{cases}
\arcsin^2\big(1/\sqrt{\tau_i}\big), & \hspace{0.7cm}\text{if }  \tau_i \geq 1,\\    
-\frac14\big[\ln\big(\frac{1+\sqrt{1-\tau_i}}{1-\sqrt{1-\tau_i}}\big)-i\pi\big]^2, & \hspace{0.7cm}\text{if }  \tau_i <1,
\end{cases}\\
g(\tau_i)&= \begin{cases}
\sqrt{\tau_i-1}\arcsin^2\big(1/\sqrt{\tau_i}\big), & \text{if }  \tau_i \geq 1,\\    
-\frac{\sqrt{1-\tau_i}}2\big[\ln\big(\frac{1+\sqrt{1-\tau_i}}{1-\sqrt{1-\tau_i}}\big)-i\pi\big]^2, & \text{if }  \tau_i <1.
\end{cases}
\end{align}
The subscript of $F_s$ and  $A_s$ indicates the spin of corresponding particle in the loop, i.e. $s\!=\!1/2$ corresponds to quarks for QCD, and quarks and charged leptons for QED; whereas, $s\!=\!1$ implies charged gauge bosons, i.e. $W^\pm$.
The corresponding twin sector loop functions are straightforward to obtain from the above generic formulae by appropriately replacing the masses, couplings, etc. 

\providecommand{\href}[2]{#2}\begingroup\raggedright\endgroup

\end{document}